\documentclass[sigconf]{acmart}

\usepackage{graphicx}
\usepackage{subcaption}
\usepackage{amsmath}
\usepackage{amsthm}

\usepackage{amssymb}
\usepackage{booktabs}
\usepackage{enumitem}

\usepackage{array}
\usepackage{multirow}
\newtheorem{assumption}{Assumption}
\usepackage{pifont}
\usepackage{balance}

\AtBeginDocument{%
  }

\copyrightyear{2025}
\acmYear{2025}
\setcopyright{acmlicensed}
\acmConference[MM '25]{Proceedings of the 33rd ACM International Conference on Multimedia}{October 27--31, 2025}{Dublin, Ireland}
\acmBooktitle{Proceedings of the 33rd ACM International Conference on Multimedia (MM '25), October 27--31, 2025, Dublin, Ireland}
\acmDOI{10.1145/3746027.3755227}
\acmISBN{979-8-4007-2035-2/2025/10}

\begin{document}

\title{FFCBA: Feature-based Full-target Clean-label Backdoor Attacks}

\author{Yangxu Yin}
\orcid{0009-0002-9517-2667}
\affiliation{%
  \institution{China University of Petroleum (East
 China)}
  \city{Qingdao}
  \country{China}
}
\email{yinyangxv@163.com}

\author{Honglong Chen}
\authornote{Corresponding authors.}
\affiliation{%
  \institution{China University of Petroleum (East
 China)}
  \city{Qingdao}
  \country{China}
}
\email{chenhl@upc.edu.cn}

\author{Yudong Gao}
\affiliation{%
  \institution{China University of Petroleum (East
 China)}
  \city{Qingdao}
  \country{China}
}
\email{YudongGao0504@163.com}

\author{Peng Sun}
\affiliation{%
  \institution{Hunan University}
  \city{Changsha}
  \country{China}
}
\email{psun@hnu.edu.cn}

\author{Liantao Wu}
\affiliation{%
  \institution{East China Normal University}
  \city{Shanghai}
  \country{China}
}
\email{ltwu@sei.ecnu.edu.cn}

\author{Zhe Li}
\affiliation{%
  \institution{China University of Petroleum (East
 China)}
  \city{Qingdao}
  \country{China}
}
\email{lizhe_upc@outlook.com}

\author{Weifeng Liu}
\affiliation{%
  \institution{China University of Petroleum (East
 China)}
  \city{Qingdao}
  \country{China}
}
\email{liuwf@upc.edu.cn}

\renewcommand{\shortauthors}{Yangxu Yin et al.}

\begin{abstract}
Backdoor attacks pose a significant threat to deep neural networks, as backdoored models would misclassify poisoned samples with specific triggers into target classes while maintaining normal performance on clean samples. 
Among these, multi-target backdoor attacks can simultaneously target multiple classes.
However, existing multi-target backdoor attacks all follow the dirty-label paradigm, where poisoned samples are mislabeled, and most of them require an extremely high poisoning rate.
This makes them easily detectable by manual inspection.
In contrast, clean-label attacks are more stealthy, as they avoid modifying the labels of poisoned samples. 
However, they generally struggle to achieve stable and satisfactory attack performance and often fail to scale effectively to multi-target attacks.
To address this issue, we propose the Feature-based Full-target Clean-label Backdoor Attacks (FFCBA) which consists of two paradigms: Feature-Spanning Backdoor Attacks (FSBA) and Feature-Migrating Backdoor Attacks (FMBA). 
FSBA leverages class-conditional autoencoders to generate noise triggers that align perturbed in-class samples with the original category's features, ensuring the effectiveness, intra-class consistency, inter-class specificity and natural-feature correlation of triggers. 
While FSBA supports swift and efficient attacks, its cross-model attack capability is relatively weak.
FMBA employs a two-stage class-conditional autoencoder training process that alternates between using out-of-class samples and in-class samples. 
This allows FMBA to generate triggers with strong target-class features, making it highly effective for cross-model attacks.
We conduct experiments on multiple datasets and models, the results show that FFCBA achieves outstanding attack performance and maintains desirable robustness against the state-of-the-art backdoor defenses.
Our source code is available at (\url{https://github.com/YangxvYin/FFCBA_code}).

\end{abstract}




\maketitle

\section{Introduction}
Deep Neural Networks (DNNs) are widely used due to their high performance. However, the lack of transparency and interpretability of DNNs makes them highly vulnerable to backdoor attacks \cite{r39,r40,r41,r8,r19}. These attacks occur when an adversary embeds a backdoor into the model during training by manipulating dataset \cite{r5,r6,r7} or altering model parameters \cite{r8,r9,r10}. Consequently, the model behaves normally on clean samples but misclassifies samples with triggers into target classes during inference. 
While backdoor attacks on image classification tasks have grown increasingly sophisticated, the majority of these attacks are single-target, meaning they can only designate one specific class as the target. 
In contrast, multi-target backdoor attacks \cite{r11,r12,r13,r14} can target multiple or even all classes (i.e. full-target backdoor attacks) simultaneously and each target class is mapped to a specific trigger injection paradigm. This enables attackers to flexibly control the classification of poisoned samples into any desired predefined target class during inference, which means attackers can switch targets for maximum  benefit. For instance, autonomous vehicles plan their routes in real-time based on roadside signs. With multi-target backdoor attacks, attackers can control vehicles to drive along any desired path by slightly modifying the roadside signs. Therefore, multi-target backdoor attacks with powerful payloads pose a significant threat to deep models, attracting much attention from both academia and industry.

\begin{figure}[!t]
    \centering
    \includegraphics[width=0.35\textwidth]{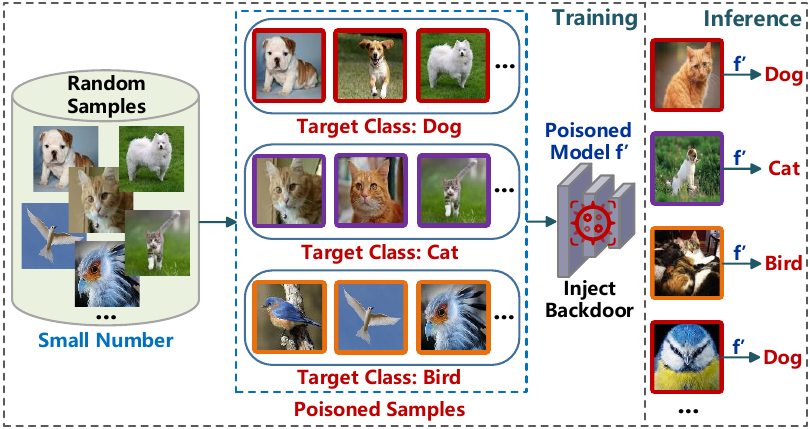}
    \caption{Schematic of low-poisoning-rate clean-label multi-target backdoor attack, where poisoned samples retain the same labels as their clean counterparts. }
    \Description{Schematic of low-poisoning-rate clean-label multi-target backdoor attacks during training and inference phases.}
    \label{fig0:env}
\end{figure}

Existing multi-target backdoor attacks have two critical flaws. One is needing to modify the labels of poisoned samples. The other is that most of them require a relatively high poisoning rate. Consequently, each target class has many poisoned samples not truly belonging to this category, making the attack highly vulnerable to detection.
Therefore, maintaining original labels with a clean-label attack paradigm \cite{r1} while keeping a low poisoning rate is essential for successful attacks.
However, implementing low-poisoning-rate clean-label multi-target or even full-target backdoor attacks, as depicted in Figure \ref{fig0:env}, poses two major challenges.

(1) \textit{Existing dirty-label multi-target backdoor attacks are inapplicable to clean-label constraint and ineffective in reducing poisoning rates.} 
Specifically, the feature strength of dirty-label triggers is typically much weaker than the natural features of the samples. Therefore, in clean-label attacks, the model is unlikely to learn these weak trigger features and will instead focus on natural ones, causing the trigger to lose effectiveness.
Moreover, most dirty-label triggers' features have low correlation with natural ones. Thus, the model needs numerous poisoned samples to identify trigger patterns, making it extremely difficult to reduce the poisoning rate.
 
(2) \textit{Existing clean-label single-target attacks fail to achieve stable and satisfactory results and are hard to extend to multi-target ones.} 
Specifically, these attacks typically employ high-intensity noise triggers, such as adversarial noise \cite{r3,r4} and noise with strong features \cite{r2}. This results in randomness in the noise triggers generated for different samples, making it difficult to ensure strong feature consistency across all poisoned samples. Consequently, the attack effect under clean-label constraint is weakened, and the attack success rate (ASR) cannot remain stable above 99\% in various datasets and models. 
Additionally, similarity in the form of noise triggers complicates the design of class-specific trigger injection paradigms that ensure trigger specificity across different classes. This significantly limits the extension of attacks to multiple targets.
We extend the outstanding clean-label attacks, Narcissus \cite{r34} and COMBAT \cite{r36}, to multi-target attacks using different seeds. The resulting average ASR across multiple targets is only 11.13\% and 15.3\%, respectively, validating the correctness of our analysis.

Therefore, we must ensure the trigger's effectiveness, intra-class consistency, and inter-class specificity to achieve clean-label multi-target attacks with good effects. Also, we need to enhance the trigger's natural-feature correlation to reduce the poisoning rate.
To meet these essential properties, we design triggers that both obscure the natural features of clean samples and guide poisoned samples to exhibit the characteristics of the target class.

Based on the above ideas, we first propose the Feature-Spanning full-target clean-label Backdoor Attack (FSBA). Specifically, FSBA employs a carefully trained class-conditional autoencoder for the attack.
When training the class-conditional autoencoder, we first overlay mid-high-frequency perturbations on samples of each class using secondary discrete wavelet transform (S-DWT). Subsequently, we use the perturbed samples along with the original class one-hot vectors as inputs to the class-conditional autoencoder. We then train it to output noise triggers that can cause the perturbed samples to re-cluster within their original feature clusters in the proxy model. This means trigger features have higher intensity than both perturbations and natural features, ensuring effectiveness and showing intra-class consistency, inter-class specificity, and natural-feature correlation at the feature level.
During the backdoor injection phase, we use clean samples and their original class vectors to generate poisoned samples for the attack. During inference, the output of the poisoned model will be consistent with the class vectors used to generate poisoned samples. 
For FSBA's class-conditional autoencoder, each class's trigger-generation paradigm is trained only based on data from that category. This allows FSBA to perform rapid and efficient attacks. However, the limited single-category data weakens the class-conditional autoencoder's generalization ability and noise features, reducing its cross-model attack capability.

Therefore, we further propose the Feature-Migrating full-target clean-label Backdoor Attack (FMBA). FMBA follows the same attack process as FSBA but employs a different training paradigm for the class-conditional autoencoder. We design a new two-stage process to train the class-conditional autoencoder. First, we train the noise trigger to migrate the features of samples outside the target class into the feature cluster of the target class, and illustrate this idea theoretically. Specifically, we use Neural Tangent Kernel theory (NTK) \cite{r30} to show that when data approaches a uniform distribution, the feature strengths of each class become similar. This indicates that when the noise trigger obscures the features of samples outside the target class, it can also obscure the features of the target class. Thus we use abundant out-of-class samples to enhance FMBA's cross-model attack capability. Second, we fine-tune the noise trigger using samples from target class to ensure a reasonable distribution of poisoned features during the attack phase, guaranteeing the four essential properties of the trigger.

\begin{figure*}[!t]
    \centering
    \includegraphics[width=0.85\textwidth]{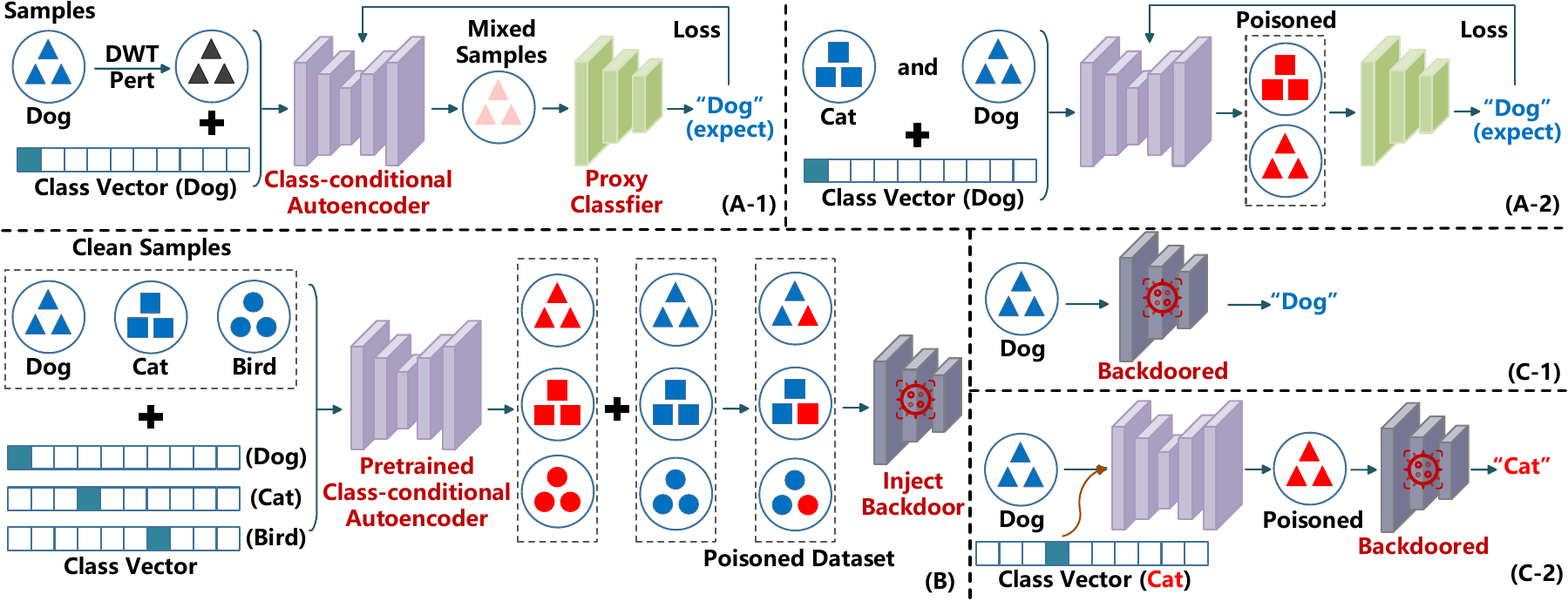}
    \caption{The attack process of FFCBA, where A-1 and A-2 represent the class-conditional autoencoder training processes of FSBA and FMBA, respectively; B is the process of injecting backdoors into the victim classification model (common to both FSBA and FMBA); C-1 and C-2 describe the outputs of the poisoned model for clean samples and poisoned samples, respectively.}
    \Description{The complete attack process of FFCBA.}
    \label{fig1:env}
\end{figure*}

In summary, when the victim model type is known, FSBA can execute rapid and efficient attacks; when the victim model type is unknown, FMBA can accomplish attacks with excellent cross-model attack capabilities. They are complementary in application scenarios and jointly constitute Feature-based Full-target Clean-label Backdoor Attacks (FFCBA). 
Our contributions are as follows:
\begin{itemize}
    \item FFCBA consists of FSBA and FMBA enabling low-poisoning clean-label full-target attacks. FSBA is more efficient, while FMBA has excellent cross-model attack capability.
    \item FFCBA enables triggers to obscure natural features and exhibit robust features corresponding to the target class, ensuring the effectiveness, intra-class consistency, inter-class specificity, and natural-feature correlation.
    \item Experiments show FFCBA delivers strong attacks with negligible benign accuracy impact and robust defense resistance.
\end{itemize}


\section{Related Work}

\textbf{Dirty-label Backdoor Attacks.}
BadNets \cite{r6} first unveiled the backdoor attack chapter, with subsequent studies \cite{r15,r16,r5,r17,r18,r29} enhancing stealth and potency. However, these efforts concentrated on single-target attacks, resulting in limited attack capabilities. Works like One-to-N \cite{r11}, Marksman \cite{r13}, and Universal Backdoor \cite{r14} break this limitation achieving multi-target attacks. However, they all operate under dirty label conditions and, except for Universal Backdoor \cite{r14}, require extremely high poisoning rates. Thus, they struggle to evade human detection, often leading to failure. Addressing the challenge of conducting low-poisoning-rate clean-label multi-target attacks is an urgent issue that requires resolution.

\noindent\textbf{Clean-label Backdoor Attacks.}
Label-consistent attack \cite{r1} first achieved clean-label backdoor attacks and conducted an in-depth analysis of the reasons for previous attack failures. It laid the foundational argument for subsequent clean-label backdoor attacks that use noise to interfere with clean features. Works such as Invisible Poison \cite{r2}, CSSBA \cite{r3}, and Poison Frogs \cite{r4} perform clean-label backdoor attacks using adversarial or strong feature noise. Besides, Narcissus \cite{r34} uses noise to aid in feature clustering. However, these attacks fail to achieve stable and satisfactory effects. They also encounter difficulties in designing class-specific triggers, preventing their extension to a multi-target attack paradigm.

\noindent\textbf{Backdoor Defenses.}
Current backdoor defense paradigms \cite{r20,r24,r28,r25,r26,r27} can be mainly categorized into input-based defenses, model-based defenses, output-based defenses, and inhibition-based defenses.
Input-based defenses detect or disrupt triggers in samples to neutralize backdoors. 
For example, frequency-based detector \cite{r35} identifies triggers through high-frequency artifacts.
Model-based defenses detect or eliminate backdoor by analyzing or adjusting model parameters. For example, Neural Cleanse \cite{r21} obtains possible triggers by reverse engineering each class and comparing them to find potential attack targets. Fine-Pruning \cite{r22} prunes the model to destroy backdoor structures.
Output-based defenses detect the backdoor by checking model outputs. STRIP \cite{r23} perturbs potentially poisoned images with random clean images and monitors output entropy for backdoor signs.
EBBA \cite{r33} calculates output energy per class, seeking abnormal high-energy values indicative of backdoors.
Inhibition-based defenses aim to train clean models based on poisoned datasets. For example, CBD \cite{r24} learns backdoor-free models directly from contaminated datasets from a causal point of view. 
However, these defense mechanisms struggle to detect FFCBA.


\section{Threat Model}
\textbf{Capability of Attackers.}
FSBA requires knowledge of the victim model and control over the dataset, whereas FMBA only needs the dataset control. This is because FSBA's cross-model attack capability is limited: the class-conditional autoencoder only works when the proxy model shares a similar architecture with the victim model.

\noindent\textbf{Attack Modeling.} In image classification, a DNN model $ f $ is trained to map images $ X $ to classes $ C = \{c_1, c_2, \ldots, c_K\} $. FFCBA designs trigger injection paradigms $ B_t(\cdot) $ for each target class $ y_t $ based on the proxy model $f_c$ to attack $f$. 
The backdoored model $ f' $ classifies any poisoned sample $ B_t(x) $ into its target class $ y_t $, which is determined by the one-hot vector used to generate $ B_t(x) $, while preserving performance on clean samples as:
\begin{equation}
\small
\begin{split}
f'(x) = y,\quad  f'(B_t(x)) = y_t,\quad x \in X,\quad  y, y_t \in C.
\end{split}
\label{equation 1}
\end{equation}

For FFCBA's backdoor injection process, the training set consists of $ N_b $ benign samples and $ N_p $ poisoned samples. The poisoned samples are obtained by sequentially selecting $ \lfloor N_p/K \rfloor $ samples $ x_t $ from each target class $ y_t $ and applying the corresponding $ B_t(\cdot) $. The labels of the poisoned samples are maintained as the original labels. In this case, the attacked DNN model $f'(\cdot;\theta)$ will be optimized according to the following optimization process:
\begin{equation}
\small
\underset{\theta}{\min}\sum_{i=1}^{N_b}\mathcal{L}(f'(x_i;\theta),y_i)+\sum_{t=1}^{K}\sum_{j=1}^{\lfloor N_p/K \rfloor}\mathcal{L}(f'(B_t(x_t^j);\theta),y_t),
\label{equation 2}
\end{equation}
where $ x_t^j $ represents the $ j $-th sample selected from class $ y_t $ and $\mathcal{L}$ denotes the cross-entropy loss. Therefore, the model will create a mapping between each $ B_t(\cdot) $ and the target class $ y_t $.

\section{Methodology}
\subsection{Motivation}

To achieve effective clean-label multi-target backdoor attacks while maintaining a low poisoning rate, each class-specific trigger injection paradigm must satisfy four key properties:
\begin{itemize}[itemindent=0pt, left=0pt]
\item \textbf{Trigger Effectiveness: } under clean-label constraint, the model can still capture trigger features, ensuring attack effectiveness.
\item \textbf{Intra-class Consistency: } the features of triggers targeting the same class must have high consistency. If class-specific trigger features are too dispersed, the model will struggle to learn their unified characteristics, reducing backdoor performance.
\item \textbf{Inter-class Specificity: }the trigger features for different classes must have sufficient differentiation. This allows the model to establish a clear one-to-one correspondence between trigger feature and corresponding target class.
\item \textbf{Natural-feature Correlation: } triggers should be highly correlated with their target class's natural features. This allows the model to enhance the learning of trigger features during benign sample training, enabling low-poisoning-rate attacks.
\end{itemize}

To address this, we must ensure that the intensity of the trigger features exceeds that of the natural features to guarantee their effectiveness. Additionally, we should align the distribution of class-specific trigger features in the feature space with the corresponding natural features to ensure intra-class consistency, inter-class specificity and natural-feature correlation.
Drawing on this principle, we introduce FFCBA which encompasses two distinct attack paradigms: FSBA and FMBA. The complete attack process is illustrated in Figure \ref{fig1:env}, and will be detailed in the following section.

\subsection{FSBA Paradigm}
FSBA employs class-conditional autoencoders for its attacks. The input to the class-conditional autoencoder consists of one-hot category vectors and clean samples, while the output is a noise trigger. These triggers  exhibit strong features aligned with the category vector, exceeding the natural features in clean samples. By mixing these triggers with clean samples, we create poisoned samples to attack the victim model. Before training the class-conditional autoencoder, we train a proxy classification model $f_c$ using clean data. Then, we train the class-conditional autoencoder following the processes (A-1) shown in Figure \ref{fig1:env}, which includes two steps:

\begin{figure*}[!t]
    \centering
    \includegraphics[width=0.91\textwidth]{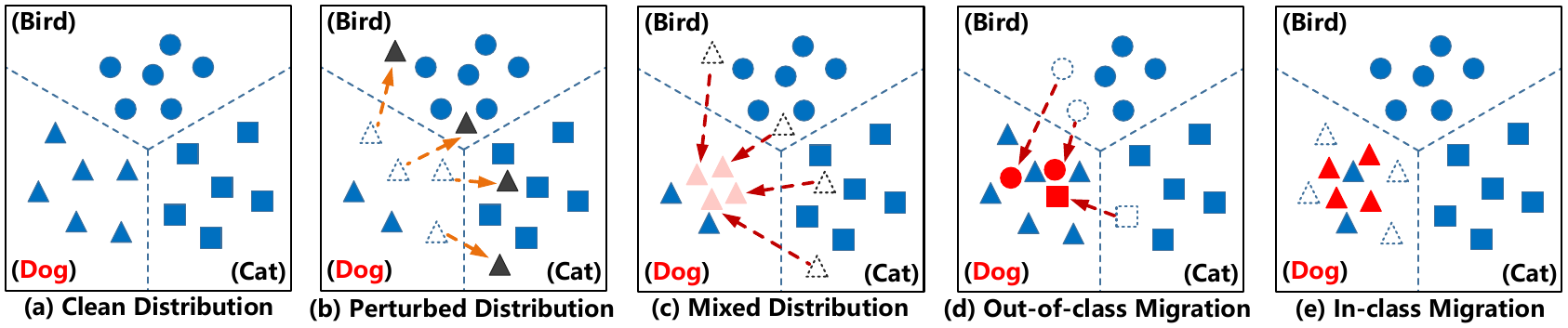}
    \caption{The feature distribution changes of target class during the training process of class-conditional autoencoders. (a) represents the distribution of clean samples in the latent space of the proxy model. (b) and (c) represent the distribution of samples after DWT perturbation and after mixing FSBA noise trigger. (d) and (e) represent the migration of out-of-class and in-class samples after mixing FMBA noise trigger, respectively. The color of samples in each state is consistent with Figure \ref{fig1:env}.}
    \Description{Schematic of feature changes in poisoned samples during the FFCBA attack process.}
    \label{fig2:env}
\end{figure*}


\textbf{Step 1: Perturbation with Natural Features.} We linearly superimpose the enhanced features from other samples onto the natural features of clean samples as perturbations. 
The feature extraction process for the samples is carried out using S-DWT and the results can be summarized as follows:
\begin{equation}
\small
\begin{split}
\text{S-DWT}(x) &= \{LL_2, HL_{1,2}, LH_{1,2}, HH_{1,2}\},\\
\Rightarrow \text{S-DWT}(x) &= \{YL, YH\},
\end{split}
\end{equation}
where $YL$ represents $LL_2$, consisting of low frequency features with the majority of the energy. $YH$ represents the remaining part, which consists of mid-high-frequency features with lower energy content. To preserve the features of clean samples, we only perturb the low-energy $ YH $ component. The specific process is:
\begin{equation}
\small
\begin{split}
&\text{S-DWT}(x_c, x_r) = \{YL_c, YH_c, YL_r, YH_r\},\\
&YH_{add} = YH_c + k\cdot YH_r, \quad (k>1)\\
&x_p = \text{IS-DWT}\{YL_c, YH_{add}\}.\quad x_c, x_r \in X,\quad x_c \neq x_r.
\end{split}
\end{equation}
We perform S-DWT on both the clean samples $x_c$ and other randomly selected samples $x_r$ from the training set. Then combine their mid-high-frequency features as $ YH_{add} $. Finally, we apply the inverse S-DWT to $ YH_{add} $ along with clean low-frequency features $ YL_c $ to obtain the perturbed samples $ x_p $. The visual effect of $ x_p $ is shown in Appendix \ref{sec:Visual}.
Although a significant amount of clean natural features is preserved, the probability of perturbed samples $ x_p $ being classified into the original category by the proxy classification model is still significantly reduced. This indicates that the intensity of the perturbation features is greater than that of the natural features, resulting in the blurring of the samples' natural characteristics as shown in Figure \ref{fig2:env} (b).

\textbf{Step 2: Feature Reconstruction Spanning the Perturbation.} 
For each category $y_k$, the class-conditional autoencoder takes perturbed samples $ x_{p,k} $ and one-hot label vector $ v_k $ as inputs, outputting noise triggers $ T_k $ matching the shape of $ x_{p,k} $.
We mix $ T_k $ with $ x_{p,k} $, denoted as $ x_{m,k} = x_{p,k} + T_k$, then input $ x_{m,k}$ into $ f_c $ and update the class-conditional autoencoder through three designed loss functions. This enables the class-conditional autoencoder to adjust features of $ x_{p,k} $, as shown in Figure \ref{fig2:env} (c).

\textbf{(1) Output Layer Loss.} 
We aim for the mixed samples $ x_{m,k} $ to be classified as their original category $ y_k $ by the proxy classification model $ f_c$.
This enables the trigger $ T_k $ to suppress the mid-high-frequency perturbations in $ x_{m,k} $, allowing $ x_{m,k} $ to exhibit the features of the original category $ y_k $.
Therefore, the trigger-feature strength $>$ perturbation strength $>$ natural-feature strength, ensuring the trigger effectiveness. Additionally, triggers for each class can exhibit robust features intrinsic to that class, guaranteeing the inter-class specificity and the natural-feature correlation.
To accomplish the aim, we formulate the output layer loss function as:
\begin{equation}
\small
\begin{split}
\mathcal{L}_{output} = \sum_{k=0}^{K}\sum_{i=0}^{n_k}\mathcal{L}(f_c(x_{p,k,i} + T_{k,i}), y_k),
\end{split}
\label{eq:output_loss}
\end{equation}
where $ K $ denotes the number of categories, $\mathcal{L}$ denotes the cross-entropy loss, $ n_k $, $x_{p,k,i}$ refer to the data volume and perturbed sample of category $ y_k $ respectively, and $T_{k,i}$ denotes sample-class-specific noise trigger.

\textbf{(2) Latent Space Loss.} Relying solely on ensuring that the mixed samples $x_{m,k}$ fall within the classification boundary of their original category $y_k$ poses challenges in meeting the intra-class consistency of the trigger. Specifically, in datasets with limited categories, the classification boundaries are often quite loose.
This results in dispersed trigger feature distributions targeting the same class, leading to insufficient consistency.
To address the issue, we calculate the centroids of the feature vector clusters for each category in the latent space of $ f_c $, denoted as $ Mean $. Here, the latent space refers to the representation of inputs in the penultimate layer of the model, denoted as $\mathcal{Z}$. Consequently, $ f_c $ can be divided into two parts: the feature extraction component $z_c : X \rightarrow \mathcal{Z}$ and the linear classification component $ l_c : \mathcal{Z} \rightarrow C$, where $f_c = z_c \circ l_c$. This means that the classification result of $ f_c $ is achieved by first applying $ z_c $, followed by $ l_c $. By constraining the distribution of the mixed sample $x_{m,k}$ in the latent space to cluster around the centroid $ Mean(y_k) $ of the original category $y_k$, we can significantly reduce the dispersion of the noise trigger features. Thus, we can derive the latent space loss as Eq. (\ref{eq:latent_loss}), where $ \mathcal{L}_1 $ denotes the L1 loss.
\begin{equation}
\small
\begin{split}
\mathcal{L}_{latent} = \sum_{k=0}^{K}\sum_{i=0}^{n_k}\mathcal{L}_1(z_c(x_{p,k,i} + T_{k,i}), Mean(y_k)).
\end{split}
\label{eq:latent_loss}
\end{equation}
It is important to note that the clustering of each class in the high-dimensional latent space is irregular. Consequently, latent space loss can only ensure that trigger features are compactly distributed. However, it cannot determine the specific class to which the trigger features belong. This implies that while intra-class consistency can be maintained, inter-class specificity cannot be guaranteed, highlighting the importance of output layer loss.

\textbf{(3) Visual Loss.}
To ensure that the poisoned samples maintain good visual quality, we impose constraints on the noise trigger from a visual perspective. Previous studies often assess the visual quality of of poisoned data using Peak Signal-to-Noise Ratio (PSNR), Structural Similarity Index Measure (SSIM), Learned Perceptual Image Patch Similarity (LPIPS), and $ l_{\infty} $ norm. However, since LPIPS can inherently affect the feature distribution of the noise trigger and frequent SSIM calculations will reduce training efficiency, we solely use PSNR and $ l_{\infty} $ norm for the visual quality constraints. Thus, we can obtain the following visual loss:
\begin{equation}
\small
\begin{split}
&\mathcal{L}_{visual} = \sum_{k=0}^{K}\sum_{i=0}^{n_k}\frac{\text{PSNR}_{\text{thresh}}
 - \text{PSNR}(x_{p,k,i}, x_{m,k,i})}{\text{PSNR}_{\text{thresh}}
},\\
&x_{m,k,i} = x_{p,k,i} + T_{k,i}, \quad s.t. \quad \|T_{k,i}\|_{\infty} \leq \epsilon,
\end{split}
\label{eq:visual_loss}
\end{equation}
where $\text{PSNR}_{\text{thresh}}$ denotes the manually set upper limit for PSNR, and $\epsilon$ represents the threshold for the $ l_{\infty} $ norm of the noise trigger. We linearly combine the three losses to form the complete loss function for training the class-conditional autoencoder, as follows:
\begin{equation}
\small
\begin{split}
\mathcal{L}_{all} = \alpha \mathcal{L}_{output}+\beta \mathcal{L}_{latent}+\gamma \mathcal{L}_{visual}.
\end{split}
\end{equation}
The necessity of each loss will be elaborated in the ablation study section.
Besides, it is important to note that in FSBA, the trigger generation paradigm for each category $ y_k $ is trained exclusively on samples from that category. The limited data volume for a single category allows the class-conditional autoencoder of FSBA to converge quickly, enabling efficient attacks. However, this can lead to weaker generalization and feature representation in noise triggers, reducing cross-model attack capabilities. As a result, FSBA requires a proxy model with an architecture similar to or on a comparable scale as the victim model to ensure attack effectiveness. To address this, we further propose FMBA based on FSBA.

\subsection{FMBA Paradigm}

FMBA uses the same attack method as FSBA with different training process of class-conditional autoencoder. To address the issues caused by insufficient data volume of single category, we choose to use out-of-class samples instead of perturbed samples. The ample supply of out-of-class samples ensures the autoencoder's generalization, and the noise trigger's ability to present strong target features significantly boosts its cross-model attack efficacy. To this end, we build upon recent studies on the neural tangent kernel (NTK) to analyze the feature strength of each category.
Specifically, we demonstrate that in some datasets, when data is uniformly distributed across categories, the feature strength of each category is similar, as follows:

\begin{assumption}\label{thm:theorem1}
For a uniformly distributed dataset and a well-trained clean model $ f $, if samples $ x_a $ and $ x_b $ from any two categories $ y_a $ and $ y_b $ are combined to obtain $ x_{\text{add}} $, denoted as $ x_{\text{add}} = x_a + x_b$, then the probability that the model $ f $ classifies $ x_{\text{add}} $ into categories $ y_a $ and $ y_b $ is approximately equal.
\end{assumption}
\begin{proof}
The brief proof of Assumption \autoref{thm:theorem1} is as follows. According to the NTK theory from previous works \cite{r30,r31,r32}, the model output of sample $x$ can be expressed as:
\begin{equation}
\begin{split}
\psi(x) = \frac{\sum_{k=0}^{K}\sum_{i=0}^{n_k}\mathcal{K}(x,x_{k,i})\cdot v_k
}{\sum_{k=0}^{K}\sum_{i=0}^{n_k}\mathcal{K}(x,x_{k,i})},
\end{split}
\end{equation}
where $ x_{k,i} $, and $ v_k $ represent the training samples, and the one-hot label vectors of category $y_k$, respectively. $K$ and $ n_k $ have the same meanings as in Eq. (\ref{eq:output_loss}).
The output is a vector with same dimension as $ v_k $, representing the probabilities of being classified into each category. Following EBBA \cite{r33}, $\mathcal{K}(x,x_{k,i}) = e^{-2\gamma||x-x_{k,i}||^2}$, $\gamma > 0$. 
The ratio of the probabilities that sample $ x_a $ being classified into category $ y_a $ and $ y_b $ by $ f $ tends toward a fixed multiple $\lambda$ as:
\begin{equation}
\begin{split}
\frac{ \sum_{i=0}^{n_a} e^{-2\gamma||x_{a,i}||^2} e^{4\gamma x_a \cdot x_{a,i}} }{ \sum_{i=0}^{n_b} e^{-2\gamma||x_{b,i}||^2} e^{4\gamma x_a \cdot x_{b,i}} } = \lambda.
\end{split}
\end{equation}
The numerator and denominator respectively represent the similarity of $ x_a $ to the samples in categories $ a $ and $ b $. 
In some datasets, such as MNIST and GTSRB, the similarity of $ x_a $ to each sample in a single category is approximately equal, and the pixel value distribution of samples in the same dataset does not have significant differences.
Moreover, $ n_a \approx n_b $. Therefore, we can derive that $e^{4\gamma x_a \cdot x_{a,i}} \approx \lambda e^{4\gamma x_a \cdot x_{b,i}}$. Similarly, for $x_b$ we have $e^{4\gamma x_b \cdot x_{b,i}} \approx \lambda e^{4\gamma x_b \cdot x_{a,i}}$. Thus we can conclude that the ratio of the probabilities of $ x_{\text{add}}$ being classified into $ y_a $ and $ y_b $ is approximately 1, as:
\begin{equation}
\begin{split}
\frac{ \sum_{i=0}^{n_a} e^{-2\gamma||x_{a,i}||^2} e^{4\gamma x_a \cdot x_{a,i}} e^{4\gamma x_b \cdot x_{a,i}} }{ \sum_{i=0}^{n_b} e^{-2\gamma||x_{b,i}||^2} e^{4\gamma x_a \cdot x_{b,i}} e^{4\gamma x_b \cdot x_{b,i}} } \approx 1.
\end{split}
\end{equation}
This suggests that the feature intensity of samples from any two categories is roughly equal; otherwise, $ f $ would classify $ x_{add} $ with high confidence into the category with stronger features. A detailed proof is provided in Appendix \ref{sec:proof}.
\end{proof}

If the noise triggers are potent enough to obscure the natural features of non-target samples while presenting the features of target class, given the similar feature intensity across different categories, these triggers should also effectively conceal the natural features of the target samples and reconstruct more robust target features, meeting the four desired properties. Based on this intuition, we design the following two-stage training process for the class-conditional autoencoder, as shown in (A-2) of Figure \ref{fig1:env}:

\textbf{Step 1: Out-of-class Feature Migration.} For each target class $y_k$, we want the corresponding noise trigger $T_k$ to make out-of-class samples $x_k^{out}$ exhibit the features of category $y_k$.
Therefore, we use $x_k^{out}$ and the target class one-hot vector $ v_k $ as the input to the class-conditional autoencoder. Similarly, we use output layer loss, latent space loss, and visual loss to constrain the output noise triggers $T_k$. Then the complete loss function is as follows:
\begin{equation}
\small
\begin{split}
&\mathcal{L}_{all} = \alpha \mathcal{L}_{output}+\beta \mathcal{L}_{latent}+\gamma \mathcal{L}_{visual},\\
&\mathcal{L}_{output} = \sum_{k=0}^{K}\sum_{i=0}^{N-n_k}\mathcal{L}(f_c(x_{k,i}^{out} + T_{k,i}), y_k),\\
&\mathcal{L}_{latent} = \sum_{k=0}^{K}\sum_{i=0}^{N-n_k}\mathcal{L}_1(z_c(x_{k,i}^{out} + T_{k,i}), Mean(y_{k})),\\
&\mathcal{L}_{visual} = \sum_{k=0}^{K}\sum_{i=0}^{N-n_k}\frac{\text{PSNR}_{\text{thresh}}
 - \text{PSNR}(x_{k,i}^{out}, x_{T,k,i}^{out})}{\text{PSNR}_{\text{thresh}}
},\\
&x_{T,k,i}^{out} = x_{k,i}^{out} + T_{k,i} \quad s.t. \quad \|T_{k,i}\|_{\infty} \leq \epsilon,
\end{split}
\label{eq:out_loss}
\end{equation}
where $N$ denotes the total data volume, $ x_{k,i}^{out} $ represents each sample outside of category $ y_k $, $ T_{k,i} $ represents the noise trigger corresponding to $ x_{k,i}^{out} $ and category $ y_k $. After above constraints, when mixing noise triggers, $ x_{k,i}^{out} $ will be classified into the target class $ y_k $ and will cluster around the centroid of the feature cluster by proxy model $f_c$, as shown in Figure \ref{fig2:env} (d), while also ensuring stealthiness.

\textbf{Step 2: In-class Feature Fine-tuning.} During the attack, poisoned samples are generated from in-class samples of the target class. Thus, for each target class $y_k$, we must ensure that in-class samples $x_k$ could exhibit strong robust target class features after superimposing the corresponding noise triggers $T_k$. Therefore, we use the in-class samples $ x_k $ and the one-hot vector $ v_k $ of category $ y_k $ as inputs to the class-conditional autoencoder. We then constrain the output noise triggers $ T_k $ using output layer loss and latent space loss. In order to refine the noise trigger features more effectively without the interference of visual factors,  we omit the visual loss at this stage. The complete loss function is:
\begin{equation}
\small
\begin{split}
&\mathcal{L}_{all} = \alpha \mathcal{L}_{output}+\beta \mathcal{L}_{latent},\\
&\mathcal{L}_{output} = \sum_{k=0}^{K}\sum_{i=0}^{n_k}\mathcal{L}(f_c(x_{k,i} + T_{k,i}), y_k),\\
&\mathcal{L}_{latent} = \sum_{k=0}^{K}\sum_{i=0}^{n_k}\mathcal{L}_1(z_c(x_{k,i} + T_{k,i}), Mean(y_{k})).
\end{split}
\label{eq:in_loss}
\end{equation}
The refinement of noise trigger features ensures that poisoned samples from various categories will cluster within their respective feature clusters in the latent space as shown in Figure \ref{fig2:env} (e),  maintaining the orderliness of multiple backdoors during the victim model's training phase.

Through the two-stage learning process, FMBA is capable of generating noise triggers with strong and robust target class features, thereby possessing excellent cross-model attack capabilities. The class-conditional autoencoder trained on any pre-trained proxy classification model can be used to attack any victim models with different architectures. We present specific attack results in the experimental section. It's important to highlight that while FMBA has learned the feature learning process of adversarial attacks, it is significantly distinct from them. The efficacy of the trigger is markedly diminished without the backdoor implantation process. For example, when a class-conditional autoencoder is trained on the Resnet18 proxy model using the ImageNet100 dataset and then used to perform backdoor-free adversarial attacks on the VGG19 and Densenet121 models, the success rates are a mere 29.72\% and 27.28\%, respectively. This underscores that FMBA's effectiveness is contingent upon the presence of a backdoor injection process.


\begin{figure}[!t]
    \centering
    \includegraphics[width=0.3\textwidth]{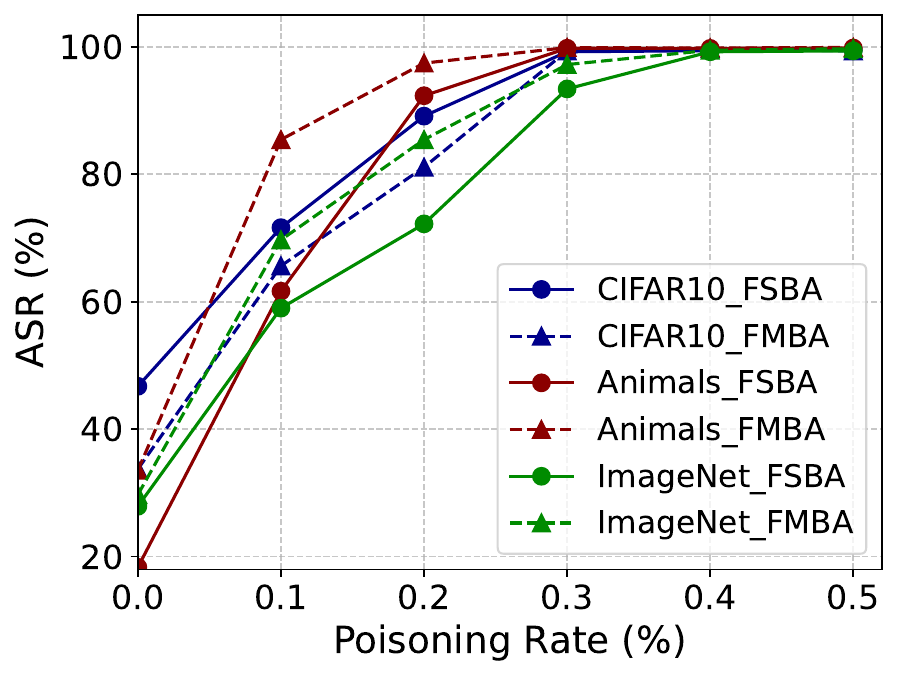}
    \caption{ASR of FFCBA across different poisoning rates with Resnet18 proxy model and VGG19 victim model. }
    \Description{The variation of FFCBA's ASR with poisoning rate.}
    \label{ASR_pr}
\end{figure}

\begin{table}[!h]
\setlength{\tabcolsep}{4.5pt}
\small
\centering
\caption{Attack properties comparison.}
\label{properties-table}
\begin{tabular}{cccccc}
\toprule
\multirow{3}{*}{Methods} & \multicolumn{5}{c}{Properties (\ding{51}/\ding{55})} \\ 
\cmidrule{2-6}
                         & \begin{tabular}[c]{@{}c@{}}Clean \\ Label\end{tabular} & \begin{tabular}[c]{@{}c@{}}Low \\ Poisoning Rate\end{tabular} & \begin{tabular}[c]{@{}c@{}}Full \\ Target\end{tabular} & \begin{tabular}[c]{@{}c@{}}Stable \\ Results\end{tabular} & \begin{tabular}[c]{@{}c@{}}Black-Box \\ Settings\end{tabular} \\ 
\midrule
FSBA                     & \ding{51} & \ding{51} & \ding{51} & \ding{51} & \ding{55} \\ 
FMBA                     & \ding{51} & \ding{51} & \ding{51} & \ding{51} & \ding{51} \\ 
One-to-N                 & \ding{55} & \ding{55} & \ding{55} & \ding{55} & \ding{51} \\ 
Marksman                 & \ding{55} & \ding{55} & \ding{51} & \ding{51} & \ding{55} \\ 
UBA                      & \ding{55} & \ding{51} & \ding{51} & \ding{51} & \ding{55} \\ 
Narcissus                & \ding{51} & \ding{51} & \ding{55} & \ding{55} & \ding{51} \\ 
COMBAT                   & \ding{51} & \ding{51} & \ding{55} & \ding{55} & \ding{51} \\ 
\bottomrule
\end{tabular}

\end{table}

\section{Evaluation}
\subsection{Experimental Setup}

\textbf{Baseline.}
Currently, no research has successfully executed clean-label multi-target attacks. Consequently, we selected three of the most advanced \textit{dirty-label} multi-target backdoor attacks—One-to-N  \cite{r11}, Marksman \cite{r13}, and Universal Backdoor Attacks \cite{r14}—as our baselines. Additionally, we have verified that state-of-the-art clean-label attack paradigms, such as Narcissus \cite{r34} and COMBAT \cite{r36}, struggle to achieve multi-target attacks by setting different seeds. Thus, we no longer include them in our baselines. Table \ref{properties-table} presents the properties of each attack paradigm. It is evident that FFCBA can achieve stable attack results under the strictest constraints.

\noindent\textbf{Dataset and Model.}
To evaluate FFCBA's performance against the baselines, we use the standard datasets for backdoor attack evaluations, including CIFAR10, Animals90, and ImageNet100. For each dataset, we employ the pre-trained Resnet18 and VGG19 to determine benign accuracy. Furthermore, we evaluate FFCBA's cross-model attack capabilities on three architecturally diverse models: Resnet50, Densenet121, and ViT\_B\_16. Details of the class-conditional autoencoder are provided in Appendix \ref{sec:autoencoder}.

\noindent\textbf{Hyperparameters.}
For training each classification model, we use the momentum Stochastic Gradient Descent (SGD) optimizer with an initial learning rate of 0.001 and a momentum of 0.9, decaying the learning rate by a factor of 0.1 every 30 epochs. For class-conditional autoencoders, we use the Adam optimizer with a learning rate of 0.0001. Besides, we set the PSNR threshold to 35 and the $ l_{\infty} $ norm constraint to 80. The parameters $ k $, $ \alpha $, $ \beta $, and $ \gamma $ are set to 1.5, 0.5, 0.3, and 0.5, respectively. For CIFAR10, Animals90, and ImageNet100, the poisoning rate is uniformly set at 0.4\%. For each baseline, we replicated its optimal experimental setup to ensure successful results.

\subsection{Effectiveness of FFCBA}

\textbf{Attack Effectiveness.}
Table \ref{sample-table1} compares FFCBA's attack performance with baselines. FSBA and FMBA, using class-conditional autoencoders trained on the Resnet18 proxy model, target the Resnet18 victim model and execute cross-model attacks on VGG19. We outline the average ASR across all labels, Benign classification Accuracy (BA), and the Decrease Value in BA (DV) compared to the clean model.
FFCBA achieves high ASR across all labels and datasets with minimal BA impact. It significantly outperforms One-to-N in ASR while matching Marksman and Universal Backdoor effectiveness under stricter clean-label constraints.
Moreover, FFCBA maintains 0.4\% poisoning rate, substantially lower than Marksman (10\%) and One-to-N (20\%), comparable to Universal Backdoor (0.62\%).
Figure \ref{ASR_pr} shows FFCBA's ASR variation with poisoning rates. 
In addition, FFCBA's poisoned samples exhibit superior visual quality, as illustrated in Appendix \ref{sec:Visual}.
Table \ref{sample-table-single} presents FFCBA's ASR for each CIFAR10 label with Resnet18 victim model. FFCBA demonstrates robust attack performance in all categories. Given the large number of categories in Animals90 and ImageNet100, we omit ASR for each label individually.

\begin{table}[!t]
\setlength{\tabcolsep}{3.5pt}
\small
\centering
  \caption{Performance of FFCBA compared with baselines.}
\label{sample-table1}
\begin{tabular}{c|c|cc|cc|cc}
\hline
\multirow{2}{*}{Method}   & \multirow{2}{*}{\begin{tabular}[c]{@{}c@{}}Metric\\ (\%)\end{tabular}} & \multicolumn{2}{c|}{CIFAR10}          & \multicolumn{2}{c|}{Animals90}        & \multicolumn{2}{c}{ImageNet100}       \\
                          &                                                                        & \multicolumn{1}{c|}{\scriptsize  Resnet18} & \scriptsize  VGG19 & \multicolumn{1}{c|}{\scriptsize  Resnet18} & \scriptsize  VGG19 & \multicolumn{1}{c|}{\scriptsize  Resnet18} & \scriptsize  VGG19 \\ \hline
\multirow{3}{*}{FSBA}     & ASR                                                                    & \multicolumn{1}{c|}{99.75}    & 99.88 & \multicolumn{1}{c|}{99.33}    & 100.0   & \multicolumn{1}{c|}{99.96}    & 99.86 \\
                          & BA                                                                     & \multicolumn{1}{c|}{86.01}    & 90.64 & \multicolumn{1}{c|}{90.33}    & 90.44 & \multicolumn{1}{c|}{86.40}     & 88.45 \\
                          & DV                                                                     & \multicolumn{1}{c|}{1.21}     & 0.29  & \multicolumn{1}{c|}{0.02}     & 0.12  & \multicolumn{1}{c|}{1.85}     &0.99  \\ \hline
\multirow{3}{*}{FMBA}     & ASR                                                                    & \multicolumn{1}{c|}{99.93}    & 99.90  & \multicolumn{1}{c|}{99.78}    & 99.45 & \multicolumn{1}{c|}{99.38}    & 99.34 \\
                          & BA                                                                     & \multicolumn{1}{c|}{86.03}    & 90.43 & \multicolumn{1}{c|}{89.77}    & 90.55 & \multicolumn{1}{c|}{86.92}    & 89.36 \\
                          & DV                                                                     & \multicolumn{1}{c|}{1.19}     & 0.50  & \multicolumn{1}{c|}{0.58}     & 0.01  & \multicolumn{1}{c|}{1.33}     & 0.08  \\ \hline
\multirow{3}{*}{One-to-N} & ASR                                                                    & \multicolumn{1}{c|}{41.96}    & 58.42 & \multicolumn{1}{c|}{6.100}     & 10.15 & \multicolumn{1}{c|}{11.64}    & 10.67 \\
                          & BA                                                                     & \multicolumn{1}{c|}{81.72}    & 88.17 & \multicolumn{1}{c|}{80.22}    & 83.44 & \multicolumn{1}{c|}{82.22}    & 83.44 \\
                          & DV                                                                     & \multicolumn{1}{c|}{5.50}     & 2.76  & \multicolumn{1}{c|}{9.59}    & 7.12  & \multicolumn{1}{c|}{6.03}     & 6.00   \\ \hline
\multirow{3}{*}{Marksman} & ASR                                                                    & \multicolumn{1}{c|}{99.64}    & 100.0   & \multicolumn{1}{c|}{99.98}    & 99.67 & \multicolumn{1}{c|}{99.94}    & 99.48 \\
                          & BA                                                                     & \multicolumn{1}{c|}{86.66}    & 90.78 & \multicolumn{1}{c|}{89.56}    & 88.44 & \multicolumn{1}{c|}{81.08}    & 85.68 \\
                          & DV                                                                     & \multicolumn{1}{c|}{0.56}     & 0.15  & \multicolumn{1}{c|}{0.79}     & 2.12  & \multicolumn{1}{c|}{7.17}     & 3.76  \\ \hline
\multirow{3}{*}{UBA}      & ASR                                                                    & \multicolumn{1}{c|}{99.32}    & 100.0  & \multicolumn{1}{c|}{99.22}    & 99.71 & \multicolumn{1}{c|}{99.80}     & 99.90  \\
                          & BA                                                                     & \multicolumn{1}{c|}{83.86}    & 85.23 & \multicolumn{1}{c|}{88.67}    & 89.55 & \multicolumn{1}{c|}{88.01}    & 88.28 \\
                          & DV                                                                     & \multicolumn{1}{c|}{3.36}      & 5.70  & \multicolumn{1}{c|}{1.68}     & 1.01  & \multicolumn{1}{c|}{0.24}     & 1.16  \\ \hline
\end{tabular}
\end{table}

\begin{table}[!t]
\setlength{\tabcolsep}{4pt}
\renewcommand{\arraystretch}{1.1} 
\small
\centering
  \caption{Attack performance of FFCBA on CIFAR10 and Resnet18 for each target.}
\label{sample-table-single}
\begin{tabular}{cccllll}
\hline
Paradigm        & \multicolumn{1}{c}{Target}       & \multicolumn{5}{c}{ASR of each category (\%)}                                                                                  \\ \hline
\multirow{2}{*}{FSBA}  & \multicolumn{1}{c}{category 0-4} & \multicolumn{1}{c}{99.98} & \multicolumn{1}{c}{99.96} & \multicolumn{1}{c}{100.0} & \multicolumn{1}{l}{99.93} & 99.73 \\ 
                       & \multicolumn{1}{c}{category 5-9} & \multicolumn{1}{l}{99.98} & \multicolumn{1}{l}{99.36} & \multicolumn{1}{l}{99.74} & \multicolumn{1}{l}{99.14} & 99.72 \\ \hline
\multirow{2}{*}{FMBA} & \multicolumn{1}{c}{category 0-4} & \multicolumn{1}{c}{100.0} & \multicolumn{1}{c}{100.0} & \multicolumn{1}{c}{100.0} & \multicolumn{1}{c}{100.0} & 100.0 \\ 
                      & \multicolumn{1}{c}{category 5-9} & \multicolumn{1}{l}{99.98} & \multicolumn{1}{l}{99.98} & \multicolumn{1}{l}{99.98} & \multicolumn{1}{l}{99.53} & 100.0 \\ \hline
\end{tabular}
\end{table}

\begin{table}[!t]
\small
\setlength{\tabcolsep}{3.5pt}
\centering
  \caption{Cross-model attack performance of FFCBA on different datasets using Resnet18 proxy model.}
\label{sample-table2}
\begin{tabular}{c|c|ccc|ccc}
\hline
\multirow{2}{*}{Dataset}     & \multirow{2}{*}{\begin{tabular}[c]{@{}c@{}}Metric\\ (\%)\end{tabular}} & \multicolumn{3}{c|}{FSBA}                                                  & \multicolumn{3}{c}{FMBA}                                                   \\ \cline{3-8} 
       &
                                                                                              & \multicolumn{1}{c|}{\scriptsize Densenet} & \multicolumn{1}{c|}{\scriptsize VITB16}   & \scriptsize Resnet50 & \multicolumn{1}{c|}{\scriptsize Densenet} & \multicolumn{1}{c|}{\scriptsize VITB16}   & \scriptsize Resnet50 \\ \hline
\multirow{3}{*}{CIFAR10}     & ASR                                                                   & \multicolumn{1}{c|}{94.38}    & \multicolumn{1}{c|}{74.14} & 96.96    & \multicolumn{1}{c|}{99.49}    & \multicolumn{1}{c|}{99.88} & 99.26    \\
                             & BA                                                                    & \multicolumn{1}{c|}{87.22}    & \multicolumn{1}{c|}{98.03} & 88.05    & \multicolumn{1}{c|}{87.38}    & \multicolumn{1}{c|}{98.12} & 88.72    \\
                             & DV                                                                    & \multicolumn{1}{c|}{0.64}     & \multicolumn{1}{c|}{0.10}   & 0.71    & \multicolumn{1}{c|}{0.48}     & \multicolumn{1}{c|}{0.01}  & 0.04     \\ \hline
\multirow{3}{*}{\begin{tabular}[c]{@{}c@{}}Animals\\ 90\end{tabular}}   & ASR                                                                     & \multicolumn{1}{c|}{85.67}    & \multicolumn{1}{c|}{95.56} & 90.67    & \multicolumn{1}{c|}{100.0}      & \multicolumn{1}{c|}{100.0}   & 100.0     \\
                             & BA                                                                     & \multicolumn{1}{c|}{91.56}    & \multicolumn{1}{c|}{94.00}    & 91.72    & \multicolumn{1}{c|}{93.88}    & \multicolumn{1}{c|}{94.44} & 94.11    \\
                             & DV                                                                     & \multicolumn{1}{c|}{2.33}     & \multicolumn{1}{c|}{0.56}  & 2.44     & \multicolumn{1}{c|}{0.01}     & \multicolumn{1}{c|}{0.12}  & 0.05     \\ \hline
\multirow{3}{*}{\begin{tabular}[c]{@{}c@{}}ImageNet\\ 100\end{tabular}} & ASR                                                                   & \multicolumn{1}{c|}{98.02}    & \multicolumn{1}{c|}{99.12} & 99.08    & \multicolumn{1}{c|}{99.46}    & \multicolumn{1}{c|}{99.22} & 99.04    \\
                             & BA                                                                     & \multicolumn{1}{c|}{90.24}    & \multicolumn{1}{c|}{92.34} & 91.12    & \multicolumn{1}{c|}{90.22}    & \multicolumn{1}{c|}{92.38} & 91.18    \\
                             & DV                                                                    & \multicolumn{1}{c|}{0.06}     & \multicolumn{1}{c|}{0.32}  & 0.10      & \multicolumn{1}{c|}{0.08}     & \multicolumn{1}{c|}{0.28}  & 0.04     \\ \hline
\end{tabular}
\end{table}

\noindent\textbf{Cross-model Attack Capability.}
We further evaluate the cross-model attack capabilities of FFCBA under various conditions.
Table \ref{sample-table2} presents FFCBA's cross-model attack performance across different datasets.
Specifically, we train class-conditional autoencoders on various datasets using the Resnet18 proxy model to attack other models with significant architectural and scale differences. 
The results indicate that FMBA maintains strong cross-model capability across all datasets while FSBA declines with increasing architectural disparity.  
Table \ref{sample-table3} presents FFCBA's cross-model attack performance across different proxy models.
Specifically, we train class-conditional autoencoders on ImageNet100 using Densenet121 and ViT\_B\_16 proxy models, which have significant structural differences, to launch attacks on various models.
The results show that FMBA maintains superior cross-model attack capabilities regardless of the proxy model used, whereas FSBA performs weaker.
Despite FSBA's cross-model limitations, Table \ref{sample-table4} confirms its effectiveness with known victim models: When proxy model matches victim type, FSBA achieves superior attacks rapidly across architectures and datasets. Therefore, with known victim model, FSBA can be deployed for quick and efficient attacks. With unknown victim model or challenging proxy training, FMBA can be employed for cross-model attacks. Their application scenarios are complementary.

\begin{table}[!t]
\setlength{\tabcolsep}{3.5pt}
\small
\centering
  \caption{Cross-model attack performance of FFCBA on ImageNet100 using different proxy models.}
\label{sample-table3}
\begin{tabular}{c|c|c|ccccc}
\hline
\multirow{2}{*}{FFCBA} & \multirow{2}{*}{\begin{tabular}[c]{@{}c@{}}Proxy\\ Model\end{tabular}} & \multirow{2}{*}{\begin{tabular}[c]{@{}c@{}}Metric\\ (\%)\end{tabular}}& \multicolumn{5}{c}{Victim Model Architecture}                                                                                             \\ \cline{4-8}
                       &                                    &                                    & \multicolumn{1}{c|}{\scriptsize Resnet18} & \multicolumn{1}{c|}{\scriptsize VGG19} & \multicolumn{1}{c|}{\scriptsize Densenet} & \multicolumn{1}{c|}{\scriptsize VITB16}   & \scriptsize Resnet50 \\ \hline
\multirow{6}{*}{FSBA}  & \multirow{3}{*}{Densenet}    & ASR                                          & \multicolumn{1}{c|}{96.92}    & \multicolumn{1}{c|}{99.48} & \multicolumn{1}{c|}{99.50}     & \multicolumn{1}{c|}{99.45} & 97.86    \\   
                       &                & BA                                                        & \multicolumn{1}{c|}{87.08}    & \multicolumn{1}{c|}{89.34} & \multicolumn{1}{c|}{89.66}    & \multicolumn{1}{c|}{92.62} & 91.16    \\
                       &                       & DV                                                 & \multicolumn{1}{c|}{1.17}     & \multicolumn{1}{c|}{0.10}   & \multicolumn{1}{c|}{0.64}     & \multicolumn{1}{c|}{0.04}  & 0.06     \\ \cline{2-8} 
                       & \multirow{3}{*}{VITB16}               & ASR                                    & \multicolumn{1}{c|}{74.10}     & \multicolumn{1}{c|}{99.46} & \multicolumn{1}{c|}{62.08}    & \multicolumn{1}{c|}{99.52} & 70.48    \\
                       &                                   & BA                                     & \multicolumn{1}{c|}{86.70}     & \multicolumn{1}{c|}{89.18} & \multicolumn{1}{c|}{90.24}    & \multicolumn{1}{c|}{92.62} & 91.20     \\
                       &                                           & DV                             & \multicolumn{1}{c|}{1.55}     & \multicolumn{1}{c|}{0.26}  & \multicolumn{1}{c|}{0.06}     & \multicolumn{1}{c|}{0.04}  & 0.02     \\ \hline
\multirow{6}{*}{FMBA}  & \multirow{3}{*}{Densenet}                           & ASR                   & \multicolumn{1}{c|}{99.06}    & \multicolumn{1}{c|}{99.88} & \multicolumn{1}{c|}{99.82}    & \multicolumn{1}{c|}{99.72} & 99.62    \\
                       &                                             & BA                           & \multicolumn{1}{c|}{86.84}    & \multicolumn{1}{c|}{88.88} & \multicolumn{1}{c|}{89.70}     & \multicolumn{1}{c|}{92.42} & 89.60     \\
                       &                                             & DV                           & \multicolumn{1}{c|}{1.41}     & \multicolumn{1}{c|}{0.56}  & \multicolumn{1}{c|}{0.60}      & \multicolumn{1}{c|}{0.24}  & 1.62     \\ \cline{2-8} 
                       & \multirow{3}{*}{VITB16}                             & ASR                      & \multicolumn{1}{c|}{99.30}     & \multicolumn{1}{c|}{99.60}  & \multicolumn{1}{c|}{99.26}    & \multicolumn{1}{c|}{99.84} & 99.10     \\
                       &                                              & BA                          & \multicolumn{1}{c|}{85.60}     & \multicolumn{1}{c|}{87.64} & \multicolumn{1}{c|}{88.20}     & \multicolumn{1}{c|}{92.46} & 89.80     \\
                       &                                            & DV                            & \multicolumn{1}{c|}{2.65}     & \multicolumn{1}{c|}{1.80}   & \multicolumn{1}{c|}{2.10}      & \multicolumn{1}{c|}{0.20}  & 1.42     \\ \hline
\end{tabular}
\end{table}

\begin{table}[!t]
\setlength{\tabcolsep}{5.5pt}
\small
\centering
  \caption{Performance of FSBA using accurate proxy models.}
\label{sample-table4}
\begin{tabular}{c|c|ccccc}
\hline
\multirow{2}{*}{Dataset}     & \multirow{2}{*}{\begin{tabular}[c]{@{}c@{}}Metric\\ (\%)\end{tabular}} & \multicolumn{5}{c}{Model Architecture}                                                                                              \\ \cline{3-7}                  &
                             & \multicolumn{1}{c|}{\scriptsize Resnet18} & \multicolumn{1}{c|}{\scriptsize VGG19} & \multicolumn{1}{c|}{\scriptsize Densenet} & \multicolumn{1}{c|}{\scriptsize VITB16} & \scriptsize Resnet50 \\ \hline
\multirow{3}{*}{CIFAR10}     & ASR & \multicolumn{1}{c|}{99.75}    & \multicolumn{1}{c|}{99.98} & \multicolumn{1}{c|}{99.59}    & \multicolumn{1}{c|}{99.84}  & 99.09    \\
                             & BA & \multicolumn{1}{c|}{86.01}    & \multicolumn{1}{c|}{90.85} & \multicolumn{1}{c|}{87.47}    & \multicolumn{1}{c|}{98.02}  & 88.41    \\
                             & DV & \multicolumn{1}{c|}{1.21}     & \multicolumn{1}{c|}{0.08}  & \multicolumn{1}{c|}{0.39}     & \multicolumn{1}{c|}{0.11}   & 0.35     \\ \hline
\multirow{3}{*}{\begin{tabular}[c]{@{}c@{}}Animals\\ 90\end{tabular}}   & ASR & \multicolumn{1}{c|}{99.33}    & \multicolumn{1}{c|}{99.78} & \multicolumn{1}{c|}{100}      & \multicolumn{1}{c|}{99.88}  & 99.56    \\
                             & BA & \multicolumn{1}{c|}{90.33}    & \multicolumn{1}{c|}{90.45} & \multicolumn{1}{c|}{93.22}    & \multicolumn{1}{c|}{94.22}  & 93.67    \\
                             & DV & \multicolumn{1}{c|}{0.02}     & \multicolumn{1}{c|}{0.11}  & \multicolumn{1}{c|}{0.67}     & \multicolumn{1}{c|}{0.34}   & 0.49     \\ \hline
\multirow{3}{*}{\begin{tabular}[c]{@{}c@{}}ImageNet\\ 100\end{tabular}}  & ASR & \multicolumn{1}{c|}{99.96}    & \multicolumn{1}{c|}{99.94} & \multicolumn{1}{c|}{99.5}     & \multicolumn{1}{c|}{99.52}  & 99.96    \\
                             & BA & \multicolumn{1}{c|}{86.4}     & \multicolumn{1}{c|}{88.9}  & \multicolumn{1}{c|}{89.66}    & \multicolumn{1}{c|}{92.62}  & 91.06    \\
                             & DV & \multicolumn{1}{c|}{1.85}     & \multicolumn{1}{c|}{0.54}  & \multicolumn{1}{c|}{0.64}     & \multicolumn{1}{c|}{0.04}   & 0.16     \\ \hline
\end{tabular}
\end{table}

\subsection{Robustness against Backdoor Defenses}
We assess the robustness of FFCBA against popular backdoor defenses, including Fine-Pruning \cite{r22}, Neural Cleanse \cite{r21}, STRIP \cite{r23}, CBD \cite{r24}, EBBA \cite{r33}, ABL \cite{r37}, and IBD-PSC \cite{r38}. These defenses have proven to be effective against previous backdoor attacks. 

\begin{figure}[!t]
    \centering
    \begin{subfigure}{0.22\textwidth}
        \centering
        \includegraphics[width=\textwidth]{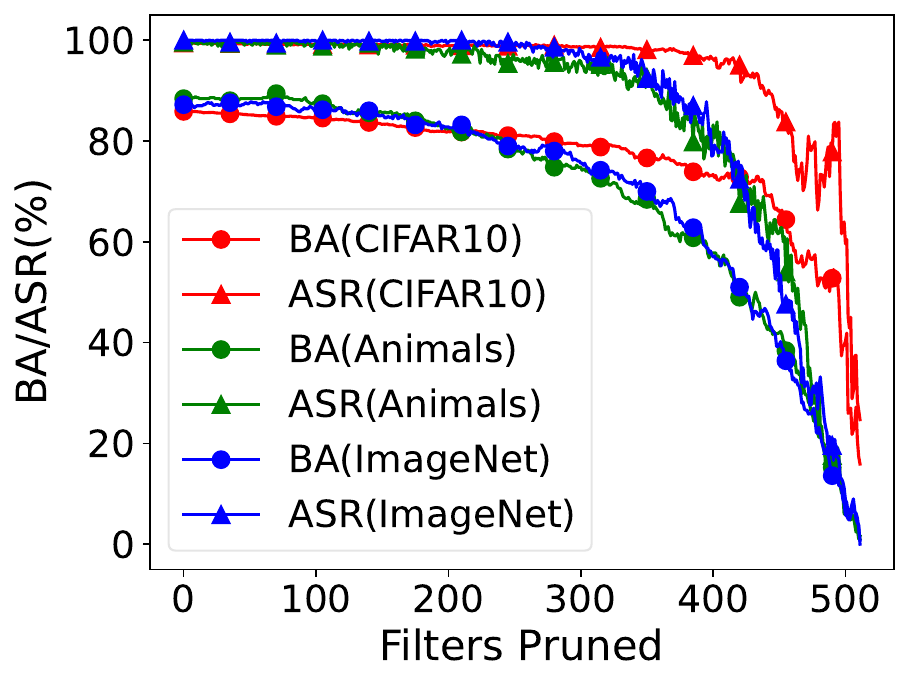}
        \caption{Fine-Pruning (FSBA)}
        \Description{Results of FSBA against Fine-Pruning.}
        \label{Fine-Pruning(FSBA)}
    \end{subfigure}
    \begin{subfigure}{0.22\textwidth}
        \centering
        \includegraphics[width=\textwidth]{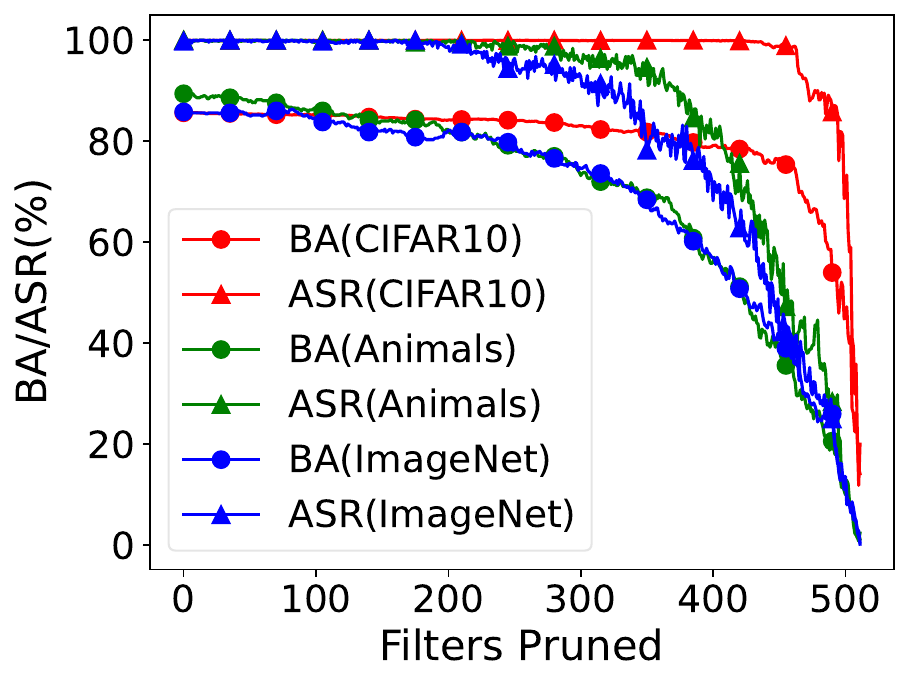}
        \caption{Fine-Pruning (FMBA)}
        \Description{Results of FMBA against Fine-Pruning.}
        \label{Fine-Pruning(FMBA)}
    \end{subfigure}
    
    \begin{subfigure}{0.22\textwidth}
        \centering
        \includegraphics[width=\textwidth]{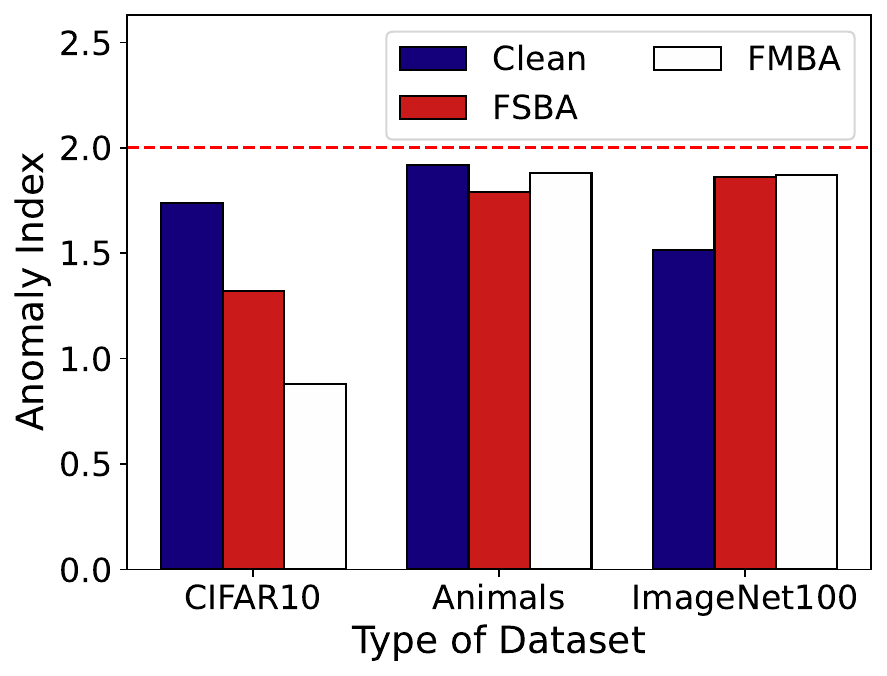}
        \caption{Neural Cleanse}
        \Description{Results of FFCBA against neural cleanse.}
        \label{defense_NC}
    \end{subfigure}
    \begin{subfigure}{0.22\textwidth} 
        \centering
        \includegraphics[width=\textwidth]{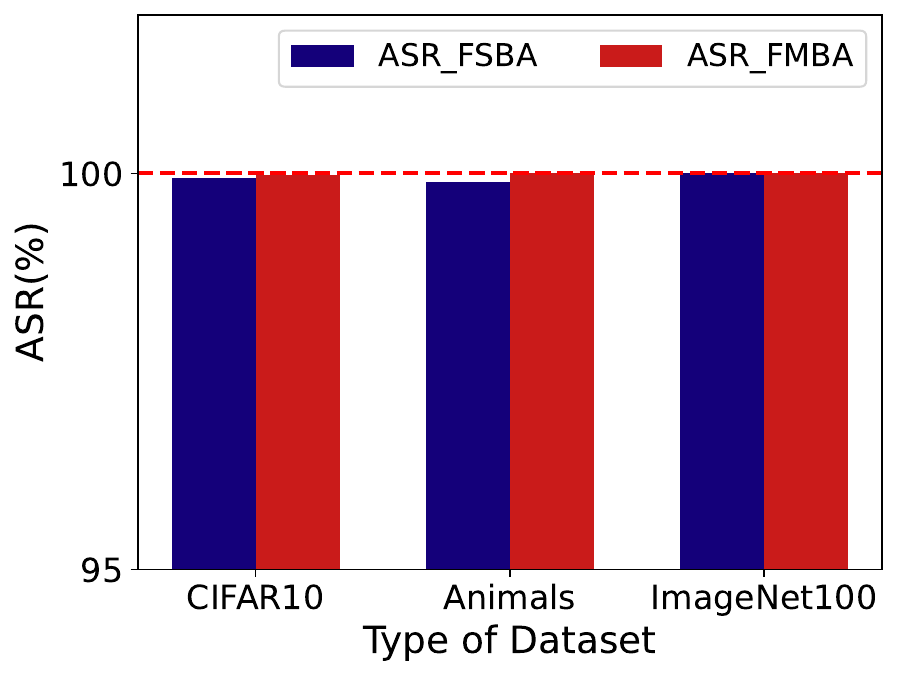}
        \caption{CBD}
        \Description{Results of FFCBA against CBD.}
        \label{defense_CBD}
    \end{subfigure}

    \begin{subfigure}{0.22\textwidth} 
        \centering
        \includegraphics[width=\textwidth]{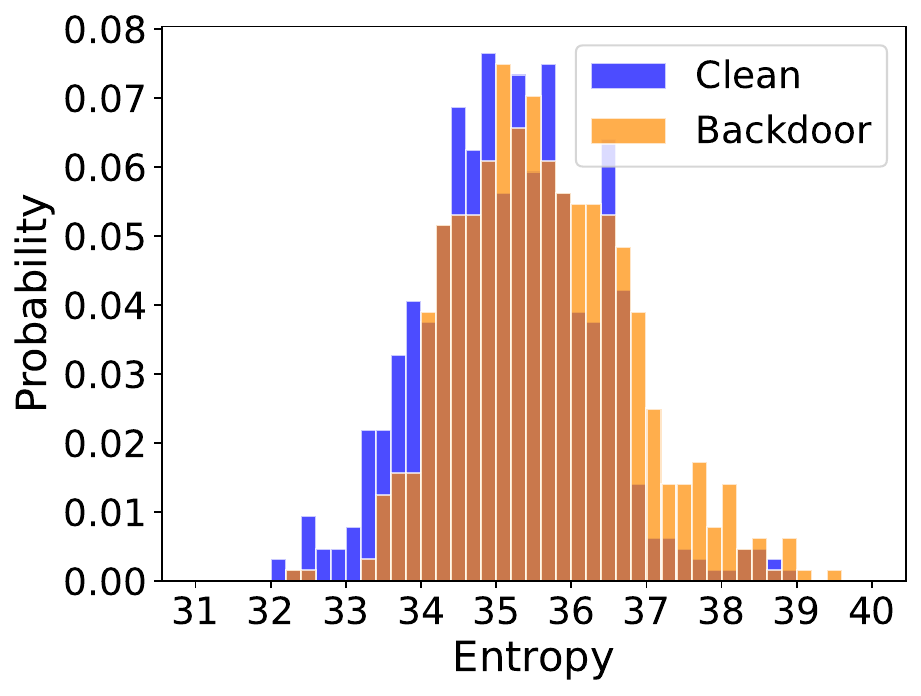}
        \caption{STRIP (FSBA)}
        \Description{Results of FSBA against STRIP.}
        \label{STRIP_Animals_FSBA}
    \end{subfigure}
    \begin{subfigure}{0.22\textwidth}
        \centering
        \includegraphics[width=\textwidth]{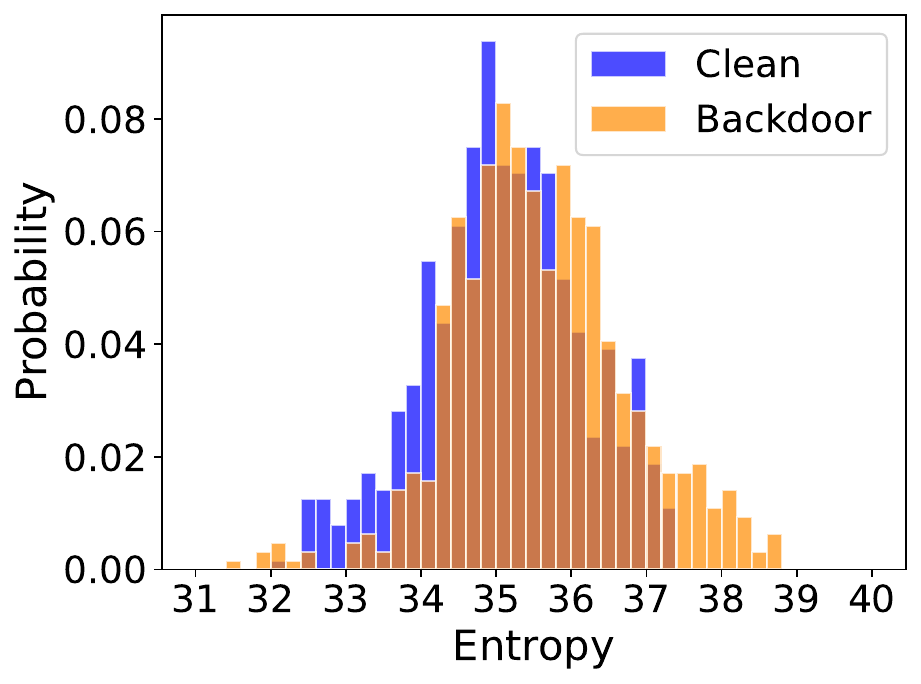}
        \caption{STRIP (FMBA)}
        \Description{Results of FMBA against STRIP.}
        \label{STRIP_Animals_FMBA}
    \end{subfigure}

    \begin{subfigure}{0.22\textwidth} 
        \centering
        \includegraphics[width=\textwidth]{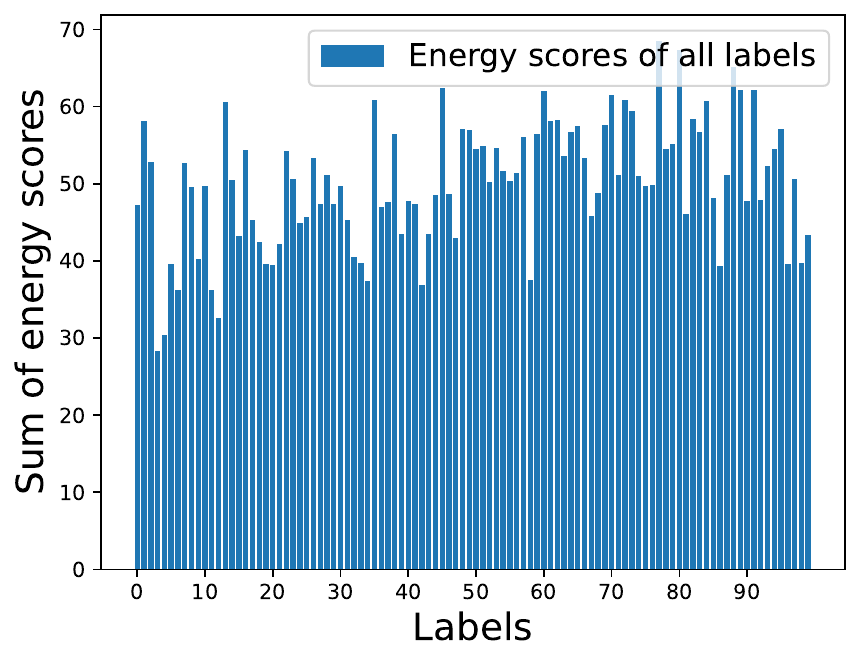}
        \caption{EBBA (FSBA)}
        \Description{Results of FSBA against EBBA.}
        \label{EBBA_ImageNet100_FSBA}
    \end{subfigure}
    \begin{subfigure}{0.22\textwidth} 
        \centering
        \includegraphics[width=\textwidth]{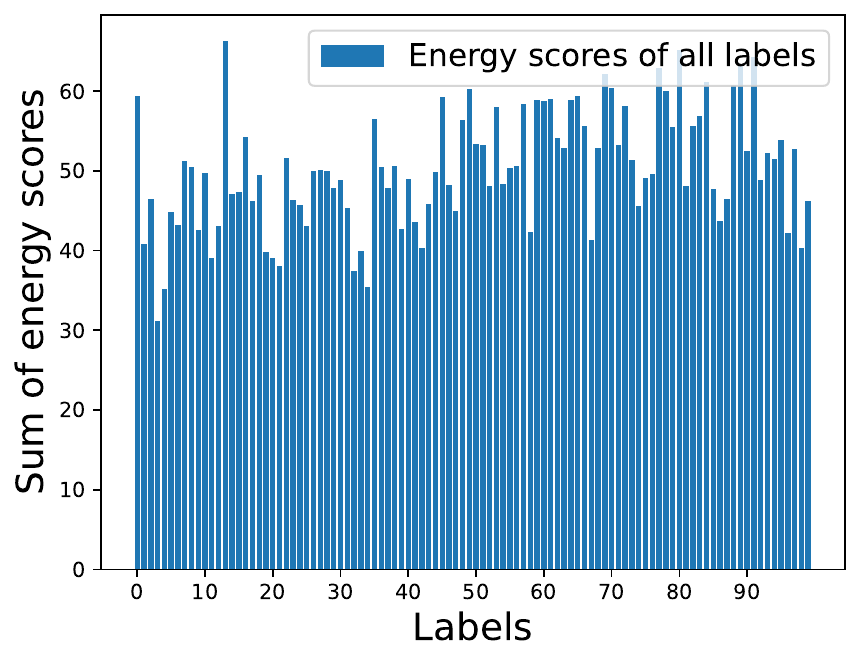}
        \caption{EBBA (FMBA)}
        \Description{Results of FMBA against EBBA.}
        \label{EBBA_ImageNet100_FMBA}
    \end{subfigure}

    \begin{subfigure}{0.22\textwidth} 
        \centering
        \includegraphics[width=\textwidth]{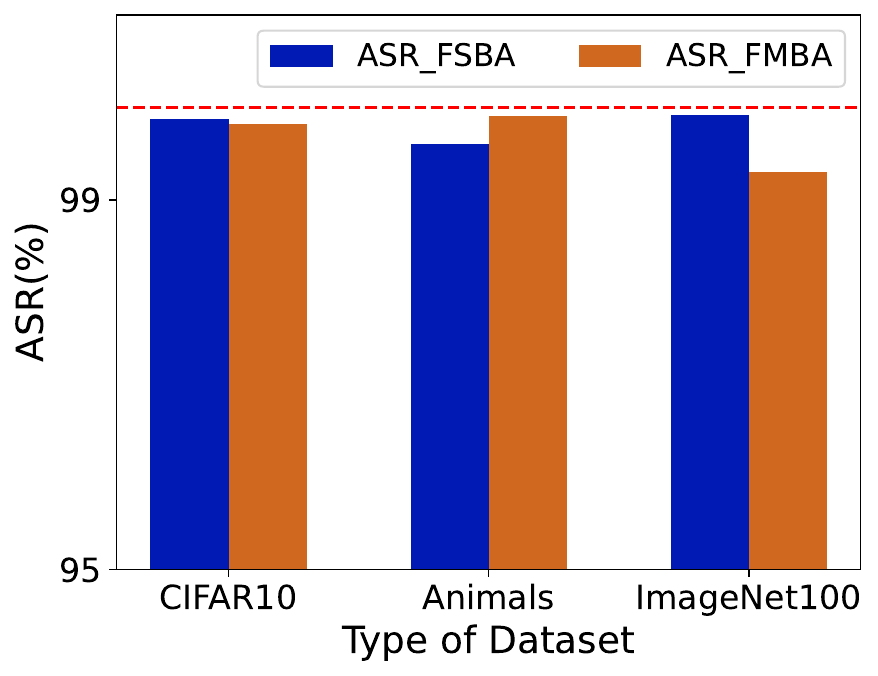}
        \caption{ABL}
        \Description{Results of FFCBA against ABL.}
        \label{ABL}
    \end{subfigure}
    \begin{subfigure}{0.22\textwidth} 
        \centering
        \includegraphics[width=\textwidth]{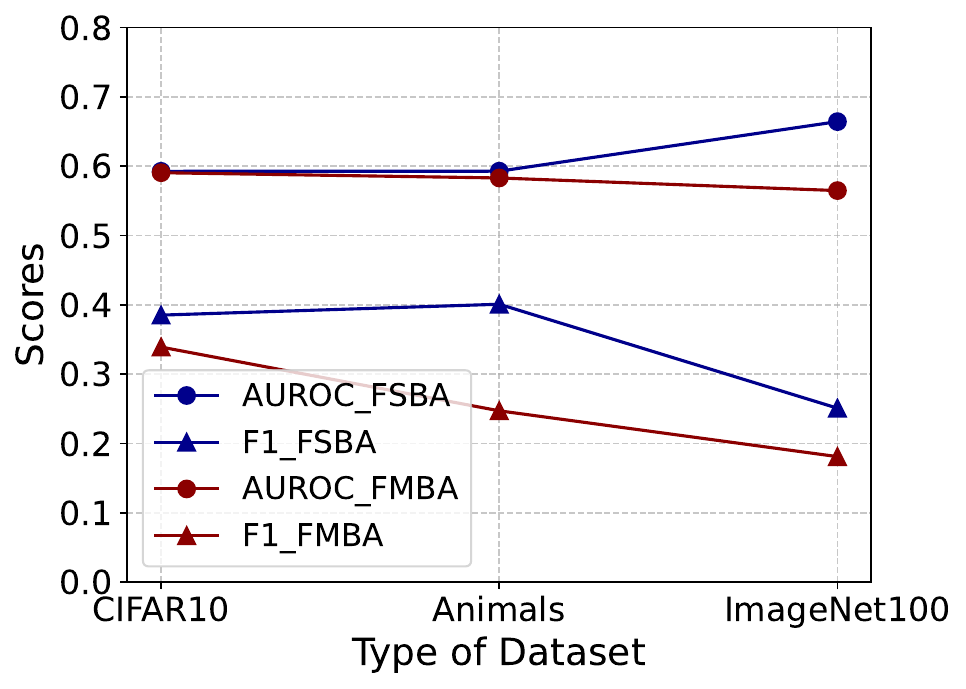}
        \caption{IBD-PSC}
        \Description{Results of FFCBA against IBD-PSC.}
        \label{IBD-PSC}
    \end{subfigure}
    
    \caption{FFCBA's performance against various defenses. }
    \Description{Results of FFCBA against various defenses.}
    \label{all_defense}
\end{figure}

\noindent\textbf{Resistance to Fine-Pruning.}
Fine-Pruning removes backdoors by pruning dormant neurons. Figures \ref{Fine-Pruning(FSBA)} and \ref{Fine-Pruning(FMBA)} demonstrate the resistance of FFCBA to Fine-Pruning. Across pruning rounds, ASR decline consistently remains below BA degradation, hence Fine-Pruning cannot defend against these two attack paradigms.

\noindent\textbf{Resistance to Neural Cleanse.}
Neural Cleanse detects backdoors by constructing reverse-engineered triggers and measuring whether anomaly metric exceed a threshold of 2. Figure \ref{defense_NC} shows both FSBA and FMBA maintain sub-threshold anomaly metrics, thus Neural Cleanse cannot defend against both attack paradigms.

\noindent\textbf{Resistance to STRIP.}
STRIP detects backdoors by perturbing inputs and identifying low-entropy outputs. Figures \ref{STRIP_Animals_FSBA} and \ref{STRIP_Animals_FMBA} show that the entropy distributions of clean and poisoned Animals90 samples for FSBA and FMBA are similar, evading detection. The results on the other two datasets are provided in Appendix \ref{sec:STRIP}.

\noindent\textbf{Resistance to CBD.}
CBD mitigates backdoors via causal learning and sample weighting on contaminated dataset. Figure \ref{defense_CBD} confirms FFCBA's resistance: both FSBA and FMBA maintain high ASR, demonstrating their robustness against CBD.

\noindent\textbf{Resistance to EBBA.}
EBBA detects backdoors by identifying labels with abnormal output energy. Figures \ref{EBBA_ImageNet100_FSBA} and \ref{EBBA_ImageNet100_FMBA} show uniform label energy distributions for FSBA and FMBA on ImageNet100 (other datasets in Appendix \ref{sec:EBBA}), hence EBBA fails in detecting FFCBA.

\noindent\textbf{Resistance to ABL.}
ABL isolates backdoor samples at the early training stage and later breaks target-class correlations to train clean models from poisoned data. Figure \ref{ABL} shows that both FSBA and FMBA maintain high ASR, indicating they can bypass ABL.

\noindent\textbf{Resistance to IBD-PSC.}
IBD-PSC detects backdoors by amplifying batch norm parameters to enhance poisoned prediction confidence consistency. Figure \ref{IBD-PSC} (threshold $T$=0.9) shows low AUROC and F1 scores for FSBA and FMBA, thus IBD-PSC fails against FFCBA.

\subsection{Ablation Study}
We conducted ablation studies on three parameters, $\alpha$, $\beta$, $\gamma$. Specifically, we trained a class-conditional autoencoder using the CIFAR10 dataset and a Resnet18 proxy model, and attacked the VGG19 model. Each parameter was varied by 0.25, while the others were fixed at their optimal values in Hyperparameters.
Table \ref{parameters-table1} and Table \ref{parameters-table2} list changes in ASR and visual metrics.
Both output layer loss and latent space loss significantly impact ASR, highlighting their importance for attack performance. While visual loss does not directly affect ASR, it greatly influences the visual metrics of poisoned samples, making it essential for maintaining visual stealthiness.


\begin{table}[!t]
\small
\centering
\caption{ASR under different parameters.}
\label{parameters-table1}
\begin{tabular}{c|ccc|ccc}
\hline
\multirow{2}{*}{Range} & \multicolumn{3}{c|}{FSBA}                                               & \multicolumn{3}{c}{FMBA}                                                \\ \cline{2-7} 
                       & \multicolumn{1}{c|}{$\alpha$} & \multicolumn{1}{c|}{$\beta$} & $\gamma$ & \multicolumn{1}{c|}{$\alpha$} & \multicolumn{1}{c|}{$\beta$} & $\gamma$ \\ \hline
0                      & \multicolumn{1}{c|}{64.68}   & \multicolumn{1}{c|}{82.70} & 99.86     & \multicolumn{1}{c|}{95.51}   & \multicolumn{1}{c|}{96.02} & 99.99    \\ \hline
0.25                   & \multicolumn{1}{c|}{95.55}   & \multicolumn{1}{c|}{97.24} & 99.84     & \multicolumn{1}{c|}{99.58}   & \multicolumn{1}{c|}{99.30} & 99.94    \\ \hline
0.5                    & \multicolumn{1}{c|}{99.80}   & \multicolumn{1}{c|}{99.78}  & 99.66    & \multicolumn{1}{c|}{99.85}   & \multicolumn{1}{c|}{99.94}   & 99.96    \\ \hline
0.75                   & \multicolumn{1}{c|}{99.23}   & \multicolumn{1}{c|}{99.85}  & 99.02    & \multicolumn{1}{c|}{99.55}   & \multicolumn{1}{c|}{99.26}   & 99.99    \\ \hline
\end{tabular}
\end{table}

\begin{table}[!t]
\small
\centering
\caption{ Visual performance under different $\gamma$.}
\label{parameters-table2}
\begin{tabular}{c|ccc|ccc}
\hline
\multirow{2}{*}{\begin{tabular}[c]{@{}c@{}}$\gamma$\\ range\end{tabular}} & \multicolumn{3}{c|}{FSBA}                                       & \multicolumn{3}{c}{FMBA}                                        \\ \cline{2-7} 
                                                                          & \multicolumn{1}{c|}{PSNR}  & \multicolumn{1}{c|}{SSIM}  & LPIPS & \multicolumn{1}{c|}{PSNR}  & \multicolumn{1}{c|}{SSIM}  & LPIPS \\ \hline
0                                                                         & \multicolumn{1}{c|}{22.43} & \multicolumn{1}{c|}{0.766} & 0.047 & \multicolumn{1}{c|}{21.29} & \multicolumn{1}{c|}{0.725} & 0.056 \\ \hline
0.25                                                                      & \multicolumn{1}{c|}{26.43} & \multicolumn{1}{c|}{0.868} & 0.021 & \multicolumn{1}{c|}{25.04} & \multicolumn{1}{c|}{0.803} & 0.041 \\ \hline
0.5                                                                       & \multicolumn{1}{c|}{30.45} & \multicolumn{1}{c|}{0.935} & 0.009 & \multicolumn{1}{c|}{30.97} & \multicolumn{1}{c|}{0.943} & 0.007 \\ \hline
0.75                                                                      & \multicolumn{1}{c|}{32.48} & \multicolumn{1}{c|}{0.957} & 0.006 & \multicolumn{1}{c|}{33.51} & \multicolumn{1}{c|}{0.976} & 0.002 \\ \hline
\end{tabular}
\end{table}

\section{Conclusion}

In this paper, we introduce FFCBA, a novel backdoor attack with two paradigms: FSBA and FMBA. FSBA uses class-conditional autoencoders to generate effective noise triggers, while FMBA extends FSBA by replacing intra-class samples with out-of-class ones, enhancing cross-model attack capabilities. Both can execute clean-label full-target backdoor attacks at low poisoning rates, effectively addressing the vulnerability of previous multi-target attack paradigms to human detection. FFCBA also demonstrates strong robustness against the state-of-the-art defenses. We demonstrate its excellent effectiveness and high defense resistance across various datasets and models.

\begin{acks}
This work was supported in part by the Shandong Provincial Taishan Scholar Program, China, under Grant tsqn202312133, Shandong Provincial Natural Science Foundation, China, under Grant ZR2022YQ61, in part by NSFC under Grants 61772551, 62111530052, 62472158, 62102337, 62202307, in part by Shandong Provincial Natural Science Foundation, China, under Grant ZR2023ZD32, Science and Technology Innovation Program of Hunan Province, under Grant 2024RC3102 and Natural Science Foundation of Hunan Province, China, under Grant 2023JJ40174.
\end{acks}

\bibliographystyle{ACM-Reference-Format}
\balance
\bibliography{main}

\appendix
\section{Appendix}

\subsection{Visual Effects of FFCBA and the Baseline}
\label{sec:Visual}
\begin{figure*}[!t]
    \centering
    \includegraphics[width=0.8\textwidth]{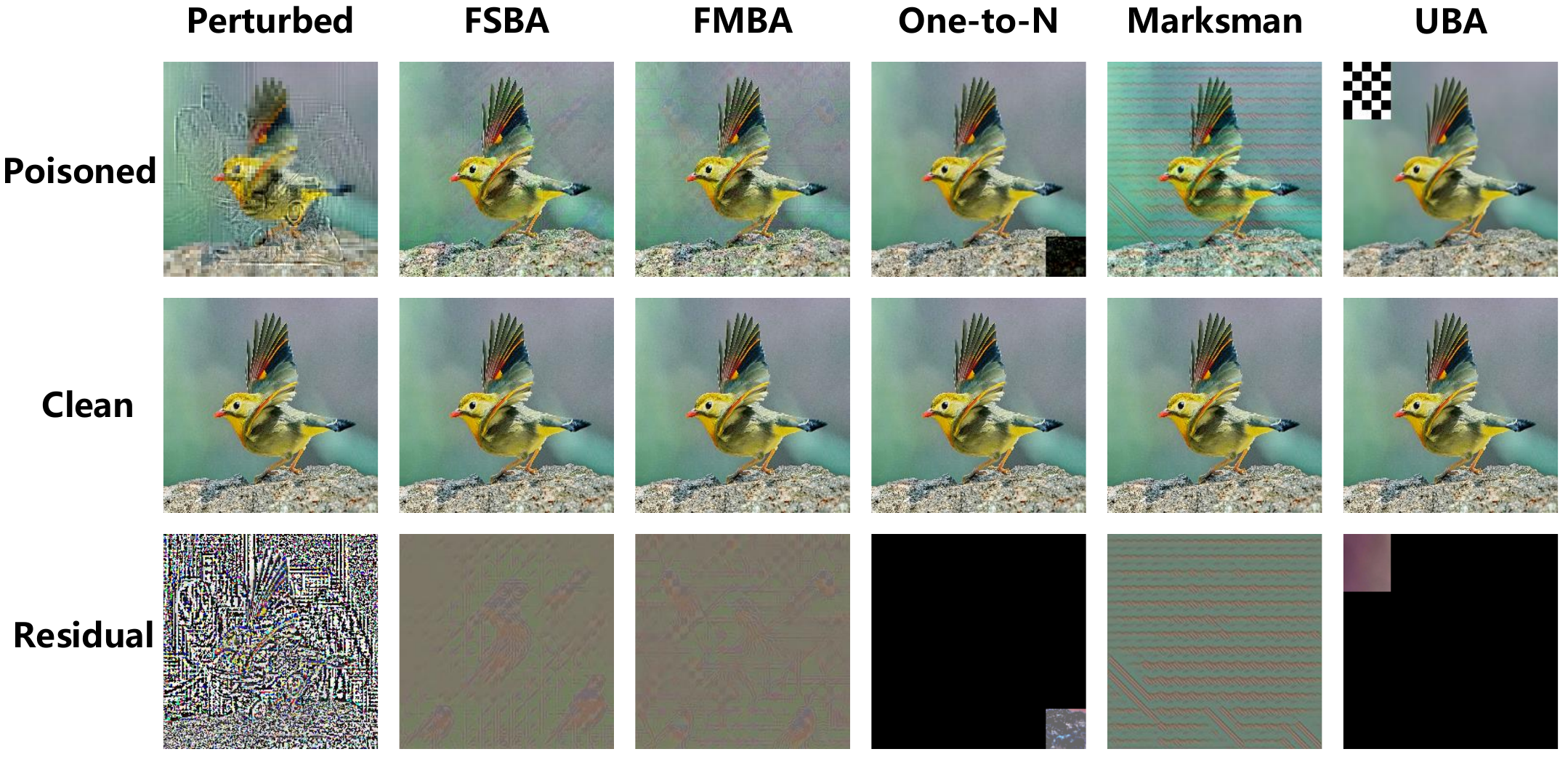}
    \caption{The visual effects of FFCBA and the baseline.
}
    \label{fig3:env}
\end{figure*}

The perturbed and poisoned samples of FFCBA, along with the baseline poisoned samples, are depicted in Figure \ref{fig3:env}. It is evident that FFCBA has good visual quality.

\begin{table*}[!t]
\normalsize
  \centering
  \caption{Class-Conditional Autoencoder Architecture.}
  \label{class-conditional autoencoder}
  \begin{tabular}{|>{\centering\arraybackslash}p{2cm}|*{5}{>{\centering\arraybackslash}p{2cm}|}}
    \toprule
    \multicolumn{1}{c}{\centering Layer} & \multicolumn{1}{c}{\centering Filters} &\multicolumn{1}{c}{\centering Filter Size} &\multicolumn{1}{c}{\centering Stride} &\multicolumn{1}{c}{\centering Padding} &\multicolumn{1}{c}{\centering Activation}\\
    \midrule
    \multicolumn{1}{c}{\centering Conv2D} & \multicolumn{1}{c}{\centering 16} &\multicolumn{1}{c}{\centering 3 $\times$ 3} &\multicolumn{1}{c}{\centering 3} &\multicolumn{1}{c}{\centering 1} &\multicolumn{1}{c}{\centering BatchNorm2D+ReLU}\\
    
    \multicolumn{1}{c}{\centering MaxPool2d} & \multicolumn{1}{c}{\centering -} &\multicolumn{1}{c}{\centering 2 $\times$ 2} &\multicolumn{1}{c}{\centering 2} &\multicolumn{1}{c}{\centering 0} &\multicolumn{1}{c}{\centering -}\\
    
    \multicolumn{1}{c}{\centering Conv2D} & \multicolumn{1}{c}{\centering 64} &\multicolumn{1}{c}{\centering 3 $\times$ 3} &\multicolumn{1}{c}{\centering 2} &\multicolumn{1}{c}{\centering 1} &\multicolumn{1}{c}{\centering BatchNorm2D+ReLU}\\
    
    \multicolumn{1}{c}{\centering MaxPool2d} & \multicolumn{1}{c}{\centering -} &\multicolumn{1}{c}{\centering 2 $\times$ 2} &\multicolumn{1}{c}{\centering 2} &\multicolumn{1}{c}{\centering 0} &\multicolumn{1}{c}{\centering -}\\
    
    \multicolumn{1}{c}{\centering ConvTranspose2D} & \multicolumn{1}{c}{\centering 128} &\multicolumn{1}{c}{\centering 3 $\times$ 3} &\multicolumn{1}{c}{\centering 2} &\multicolumn{1}{c}{\centering -} &\multicolumn{1}{c}{\centering BatchNorm2D+ReLU}\\
    
    \multicolumn{1}{c}{\centering ConvTranspose2D} & \multicolumn{1}{c}{\centering 64} &\multicolumn{1}{c}{\centering 5 $\times$ 5} &\multicolumn{1}{c}{\centering 3} &\multicolumn{1}{c}{\centering 1} &\multicolumn{1}{c}{\centering BatchNorm2D+ReLU}\\
    
    \multicolumn{1}{c}{\centering ConvTranspose2D} & \multicolumn{1}{c}{\centering 1} &\multicolumn{1}{c}{\centering 2 $\times$ 2} &\multicolumn{1}{c}{\centering 2} &\multicolumn{1}{c}{\centering 1} &\multicolumn{1}{c}{\centering BatchNorm2D+Tanh}\\
    \bottomrule
  \end{tabular}
\end{table*}

\subsection{Proof of Assumption 1}
\label{sec:proof}
\textbf{Assumption 1}
\textit{
For a uniformly distributed dataset and a well-trained clean model $ f $, if samples $ x_a $ and $ x_b $ from any two categories $ y_a $ and $ y_b $ are combined to obtain $ x_{\text{add}} $, denoted as $ x_{\text{add}} = x_a + x_b$, then the probability that the model $ f $ classifies $ x_{\text{add}} $ into categories $ y_a $ and $ y_b $ is approximately equal.
}

\begin{proof}
According to the NTK theory from previous works \cite{r30,r31,r32}, the model output of sample $x$ can be expressed as:
\begin{equation}
\begin{split}
\psi(x) = \frac{\sum_{k=0}^{K}\sum_{i=0}^{n_k}\mathcal{K}(x,x_{k,i})\cdot y_{k}
}{\sum_{k=0}^{K}\sum_{i=0}^{n_k}\mathcal{K}(x,x_{k,i})},
\end{split}
\end{equation}
where $ x_{k,i} $, and $ y_{k} $ represent the training samples, and the one-hot label vectors of category $ k $, respectively. K and $ n_k $ have the same meanings as in Eq.(5).
 The output is a vector with same dimension as $ y_{k} $, representing the probabilities of being classified into each category. Following EBBA \cite{r33}, $\mathcal{K}(x,x_{k,i}) = e^{-2\gamma||x-x_{k,i}||^2}$, $\gamma > 0$. Since the vector \( y_k \) has only the element corresponding to class \( k \) as 1, and all other positions are 0, the probability of sample \( x_a \) being classified into class \( y_a \) can be simplified as follows:
\begin{equation}
\begin{split}
\psi_a(x) = \frac{\sum_{i=0}^{n_a}\mathcal{K}(x,x_{a,i})
}{\sum_{k=0}^{K}\sum_{i=0}^{n_k}\mathcal{K}(x,x_{k,i})}.
\end{split}
\end{equation}
Similarly, the probability of \( x_a \) being classified as category \( y_b \) follows the same pattern. The probability of \( x_a \) being classified as \( y_a \) is consistently above 0.9, while for \( y_b \) it is below 0.1. Thus,  the ratio of the probabilities that $ x_a $ being classified into category $ y_a $ and $ y_b $ by $ f $ tends toward a fixed multiple $\lambda$, which can be simplified as:
\begin{equation}
\begin{split} 
&\frac{\sum_{i=0}^{n_a}\mathcal{K}(x,x_{a,i})
}{\sum_{i=0}^{n_b}\mathcal{K}(x,x_{b,i})} = \lambda\\
\Rightarrow & \frac{ e^{-2\gamma||x_{a}||^2} \sum_{i=0}^{n_a} e^{-2\gamma(||x_{a,i}||^2 - 2x_a \cdot x_{a,i})} }{ e^{-2\gamma||x_{a}||^2} \sum_{i=0}^{n_b} e^{-2\gamma(||x_{b,i}||^2 - 2x_a \cdot x_{b,i})}  } = \lambda\\
\Rightarrow & \frac{ \sum_{i=0}^{n_a} e^{-2\gamma||x_{a,i}||^2} e^{4\gamma x_a \cdot x_{a,i}} }{ \sum_{i=0}^{n_b} e^{-2\gamma||x_{b,i}||^2} e^{4\gamma x_a \cdot x_{b,i}} } = \lambda.
\end{split}
\end{equation}
The numerator and denominator respectively represent the similarity of $ x_a $ to the samples in categories $ a $ and $ b $. 
In some datasets, such as MNIST and GTSRB, the similarity of $ x_a $ to each sample in a single category is approximately equal, and the pixel value distribution of samples in the same dataset does not have significant differences.
Moreover, $ n_a \approx n_b $.
Therefore, we can derive that $e^{4\gamma x_a \cdot x_{a,i}} \approx \lambda e^{4\gamma x_a \cdot x_{b,i}}$. Similarly, for $x_b$ we have $e^{4\gamma x_b \cdot x_{b,i}} \approx \lambda e^{4\gamma x_b \cdot x_{a,i}}$. Thus we can conclude that the ratio of the probabilities of $ x_{\text{add}} = x_a + x_b $ being classified into $ y_a $ and $ y_b $ is approximately 1, as follows:
\begin{equation}
\small
\begin{split}
& \frac{ \sum_{i=0}^{n_a} e^{-2\gamma||x_{a}+x_b-x_{a,i}||^2} }{ \sum_{i=0}^{n_b} e^{-2\gamma||x_{a}+x_b-x_{b,i}||^2}  } = \frac{ e^{-2\gamma(||x_{a}||^2 + ||x_{b}||^2 + 2x_{a}\cdot x_b)}}{ e^{-2\gamma(||x_{a}||^2 + ||x_{b}||^2 + 2x_{a}\cdot x_b)} }\\
& \cdot \frac{ \sum_{i=0}^{n_a} e^{-2\gamma(||x_{a,i}||^2 - 2x_a \cdot x_{a,i} - 2x_b \cdot x_{a,i})} }{ \sum_{i=0}^{n_b} e^{-2\gamma(||x_{b,i}||^2 - 2x_a \cdot x_{b,i} - 2x_b \cdot x_{b,i})}  }\\
= &\frac{ \sum_{i=0}^{n_a} e^{-2\gamma||x_{a,i}||^2} e^{4\gamma x_a \cdot x_{a,i}} e^{4\gamma x_b \cdot x_{a,i}} }{ \sum_{i=0}^{n_b} e^{-2\gamma||x_{b,i}||^2} e^{4\gamma x_a \cdot x_{b,i}} e^{4\gamma x_b \cdot x_{b,i}} } \approx 1.
\end{split}
\end{equation}
This suggests that the feature intensity of samples from any two categories is roughly equal; otherwise, $ f $ would classify $ x_{add} $ with high confidence into the category with stronger features.

\end{proof}

\subsection{Architecture of Class-conditional Autoencoder}
\label{sec:autoencoder}
The architecture of the class-conditional autoencoder is detailed in Table \ref{class-conditional autoencoder}.

\begin{figure*}[!t]
    \centering
    \begin{subfigure}{0.32\textwidth} 
        \centering
        \includegraphics[width=\textwidth]{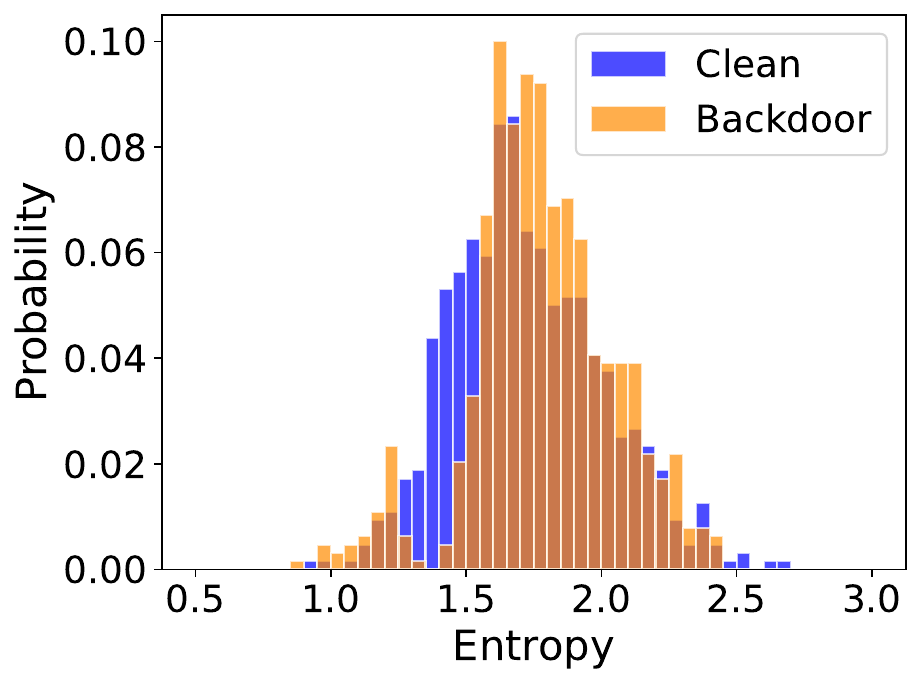}
        \caption{CIFAR10}
        \label{CIFAR10_FSBA}
    \end{subfigure}
    \begin{subfigure}{0.32\textwidth} 
        \centering
        \includegraphics[width=\textwidth]{figure/Animals_strip_FSBA.pdf}
        \caption{Animals}
        \label{Animals_FSBA}
    \end{subfigure}
    \begin{subfigure}{0.32\textwidth} 
        \centering
        \includegraphics[width=\textwidth]{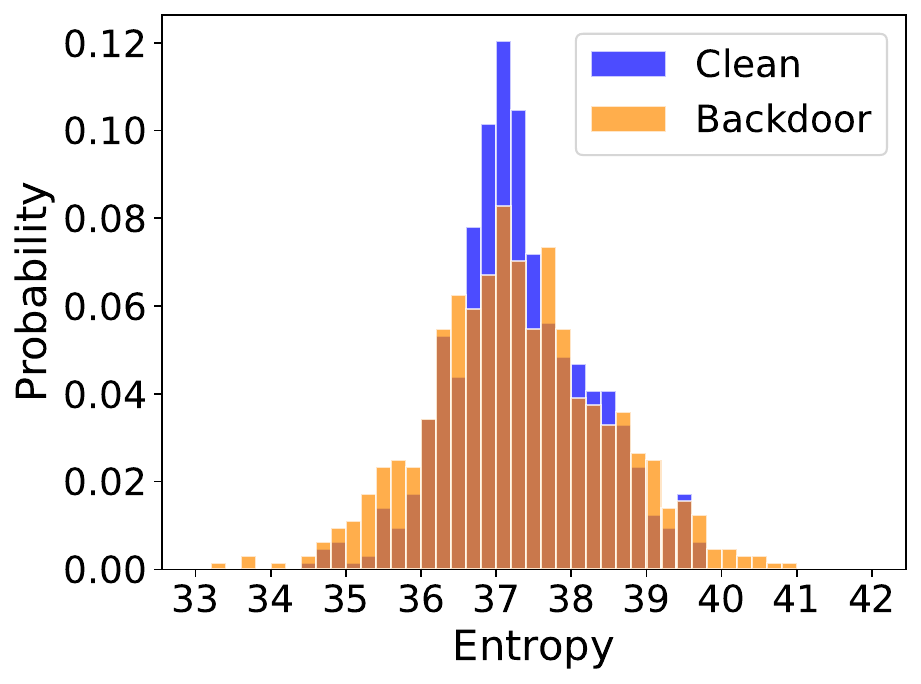}
        \caption{ImageNet100}
        \label{ImageNet100_FSBA}
    \end{subfigure}
    \caption{FSBA against STRIP.}
    \label{all_defense_strip_FSBA}
\end{figure*}

\begin{figure*}[!t]
    \centering
    \begin{subfigure}{0.32\textwidth} 
        \centering
        \includegraphics[width=\textwidth]{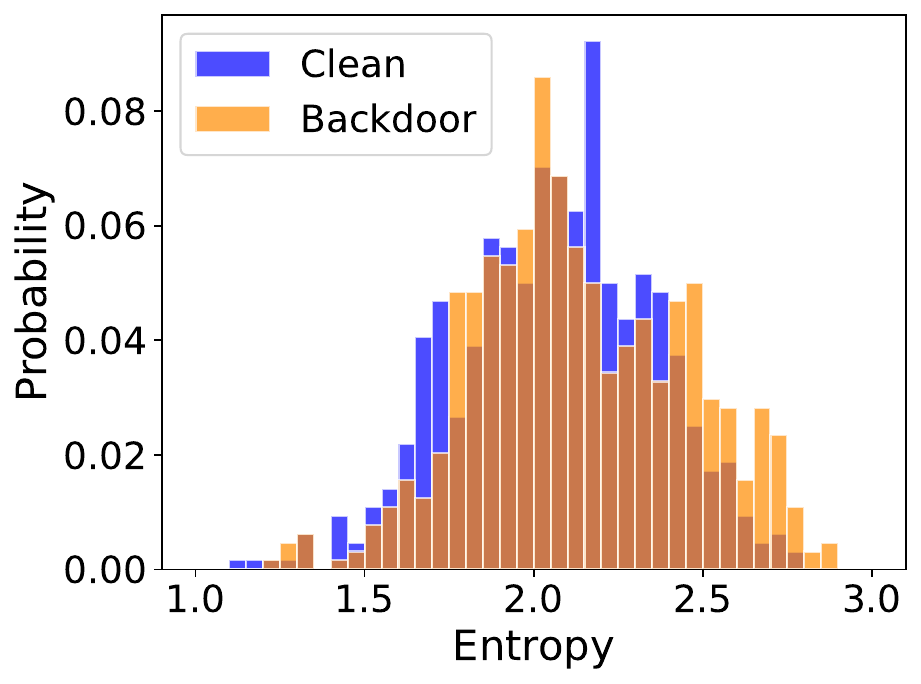}
        \caption{CIFAR10}
        \label{CIFAR10_FMBA}
    \end{subfigure}
    \begin{subfigure}{0.32\textwidth} 
        \centering
        \includegraphics[width=\textwidth]{figure/Animals_strip_FMBA.pdf}
        \caption{Animals}
        \label{Animals_FMBA}
    \end{subfigure}
    \begin{subfigure}{0.32\textwidth} 
        \centering
        \includegraphics[width=\textwidth]{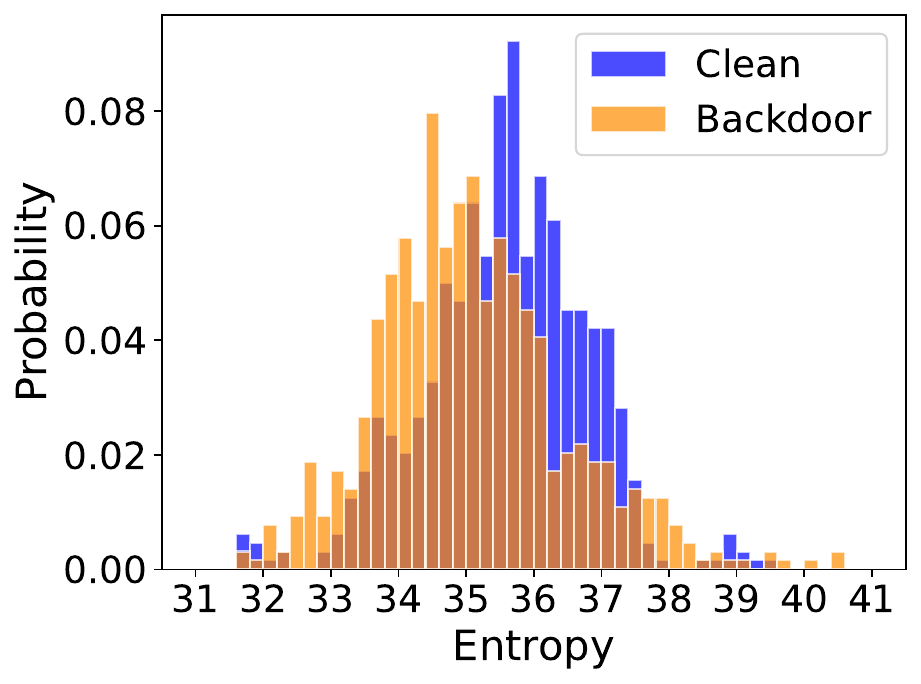}
        \caption{ImageNet100}
        \label{ImageNet100_FMBA}
    \end{subfigure}
    \caption{FMBA against STRIP.}
    \label{all_defense_strip_FMBA}
\end{figure*}

\begin{figure*}[!t]
    \centering
    \begin{subfigure}{0.32\textwidth} 
        \centering
        \includegraphics[width=\textwidth]{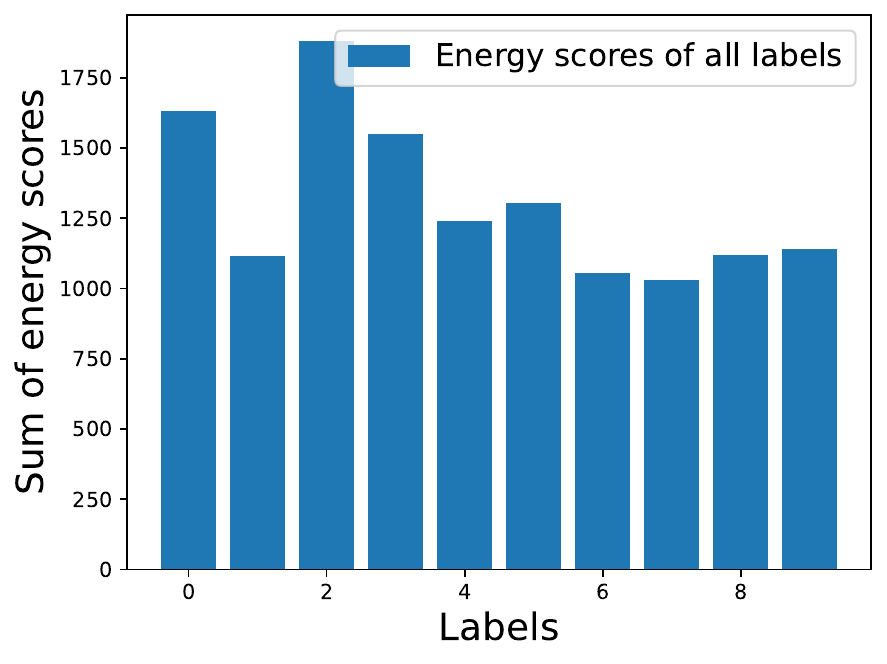}
        \caption{CIFAR10}
        \label{EBBA_CIFAR10_FSBA}
    \end{subfigure}
    \begin{subfigure}{0.32\textwidth} 
        \centering
        \includegraphics[width=\textwidth]{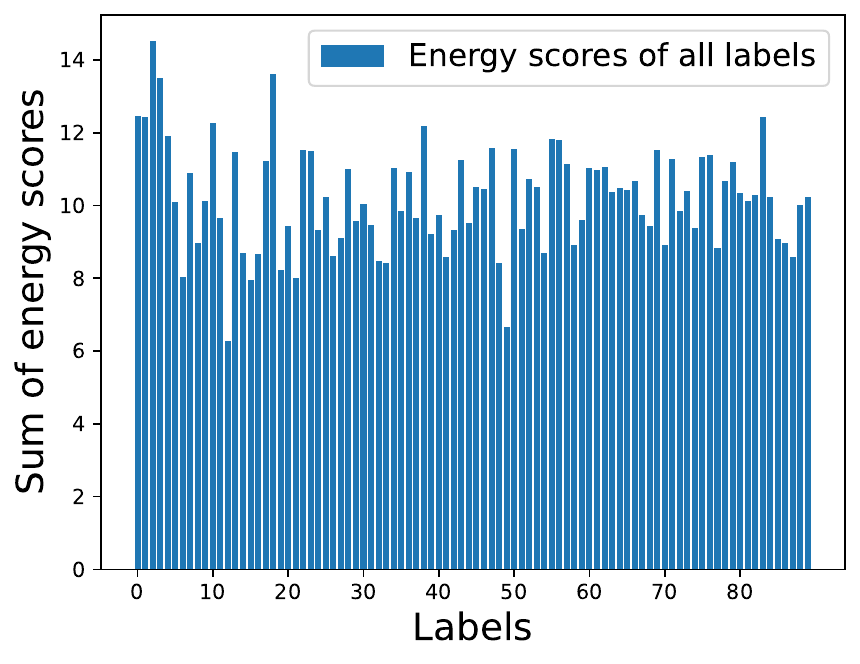}
        \caption{Animals}
        \label{EBBA_Animals_FSBA}
    \end{subfigure}
    \begin{subfigure}{0.32\textwidth} 
        \centering
        \includegraphics[width=\textwidth]{figure/EBBA_imagenet_FSBA.pdf}
        \caption{ImageNet100}
        \label{EBBA_ImageNet100_FSBA}
    \end{subfigure}
    \caption{FSBA against EBBA.}
    \label{all_defense_EBBA_FSBA}
\end{figure*}

\begin{figure*}[!t]
    \centering
    \begin{subfigure}{0.32\textwidth} 
        \centering
        \includegraphics[width=\textwidth]{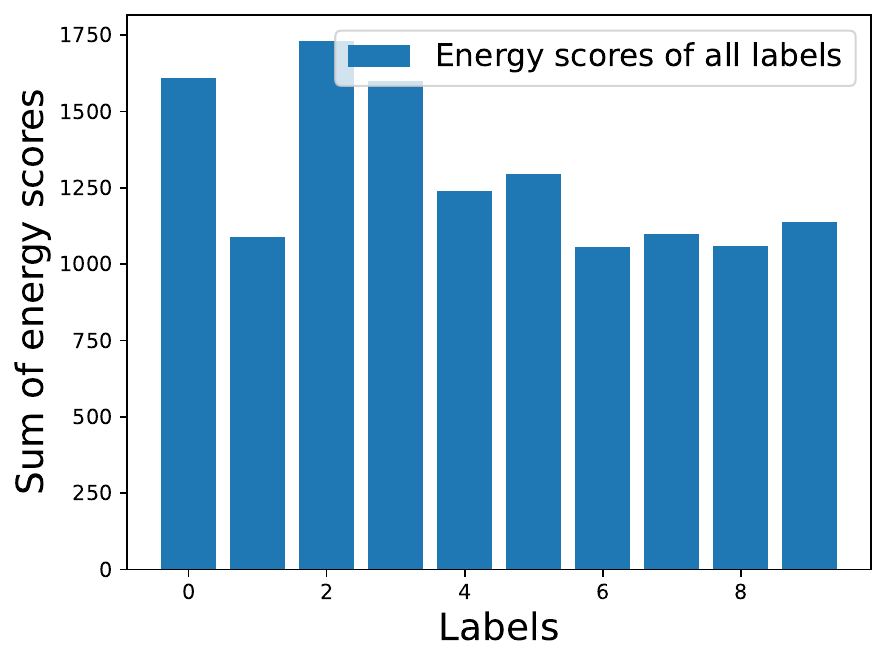}
        \caption{CIFAR10}
        \label{EBBA_CIFAR10_FMBA}
    \end{subfigure}
    \begin{subfigure}{0.32\textwidth} 
        \centering
        \includegraphics[width=\textwidth]{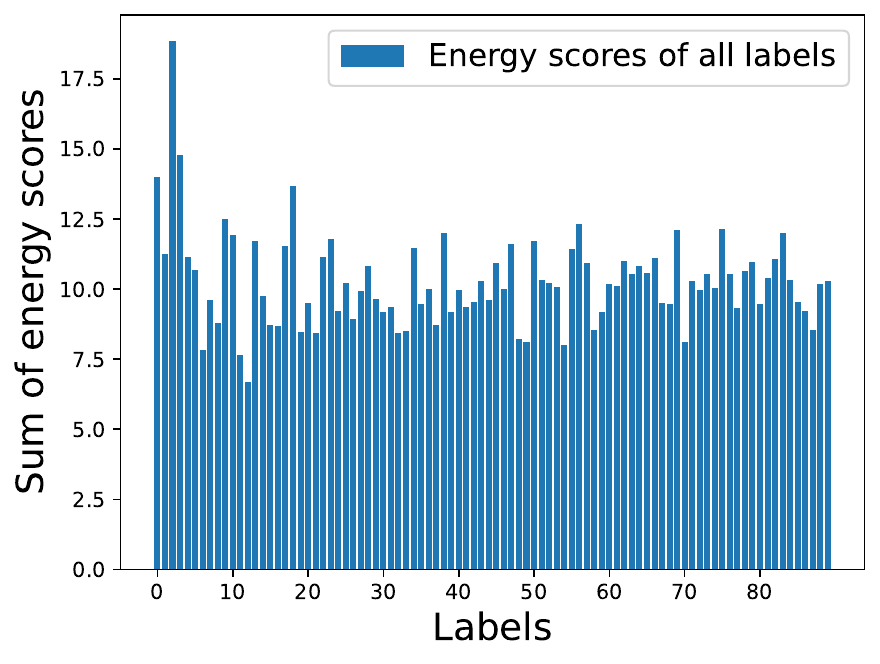}
        \caption{Animals}
        \label{EBBA_Animals_FMBA}
    \end{subfigure}
    \begin{subfigure}{0.32\textwidth} 
        \centering
        \includegraphics[width=\textwidth]{figure/EBBA_imagenet_FMBA.pdf}
        \caption{ImageNet100}
        \label{EBBA_ImageNet100_FMBA}
    \end{subfigure}
    \caption{FMBA against EBBA.}
    \label{all_defense_EBBA_FMBA}
\end{figure*}

\subsection{Resistance to STRIP}
\label{sec:STRIP}
Figure \ref{all_defense_strip_FSBA} and Figure \ref{all_defense_strip_FMBA} present the complete results of FFCBA's resistance to STRIP, showing that STRIP fails to defend against FFCBA on any dataset.

\subsection{Resistance to EBBA}
\label{sec:EBBA}
Figure \ref{all_defense_EBBA_FSBA} and Figure \ref{all_defense_EBBA_FMBA} present the complete results of FFCBA's resistance to EBBA, showing that EBBA fails to defend against FFCBA on any dataset.

\end{document}